\documentclass[11pt]{article}

\usepackage{xspace}

\newcommand{\ttbar}{\ensuremath{{t\bar{t}}}}
\newcommand{\pt}{\ensuremath{{p_{\rm T}}}}

\newcommand{\Delphes}{\textsc{Delphes}\xspace}
\newcommand{\Pythia}{\textsc{Pythia}\xspace}
\newcommand{\GeV}[1]{{#1$\,\textrm{GeV}$}\xspace}
\newcommand{\MadGraph}{\textsc{MadGraph5}\xspace}
\def\address#1{\expandafter\def\expandafter\@aabuffer\expandafter
	{\@aabuffer{\affiliationfont{#1}}\relax\par
	\vspace*{13pt}}}
\def\Hline{%
  \noalign{\ifnum0=`}\fi\hrule \@height 2\arrayrulewidth \futurelet
   \@tempa\@xhline}
\def\tablecaptionfont{\footnotesize}
\def\tablefont{\footnotesize}
\newdimen\tablewd
\long\def\tbl#1#2{%
	\parindent\z@\ignorespaces\noindent\tablecaptionfont
	\caption{#1}%
  	\par\setbox\tempbox\hbox{\tablefont #2}%
  	\tablewd\hsize\advance\tablewd-\wd\tempbox\global\divide\tablewd\tw@
	\ifdim\wd\captionbox<\wd\tempbox\centerline{\unhbox\captionbox}
	\else\leftskip\tablewd\rightskip\leftskip{\unhbox\captionbox}\par
	\fi\vskip5pt\centerline{\box\tempbox}
}%

\usepackage{authblk}
\usepackage{url}
\usepackage{hyperref}
\usepackage{ dsfont }
\usepackage{graphicx}
\usepackage{indentfirst}
\usepackage[export]{adjustbox}
\usepackage{fullpage}
\usepackage{tabularx}
\usepackage[toc,page]{appendix}
\usepackage{authblk}

\begin{document}

\title{Study of semi-boosted top quark reconstruction performance on the line shape of a $t\bar{t}$ resonance}
\author[]{J.~P\'{a}calt}
 \affil[]{\textit{Joint Laboratory of Optics of Palacky University Olomouc and Institute of Physics of Czech Academy of Sciences}}
 \affil[]{\url{josef.pacalt@upol.cz}}

\author[]{J.~Kvita}
 \affil[]{\textit{Joint Laboratory of Optics of Palacky University Olomouc and Institute of Physics of Czech Academy of Sciences}}
\affil[]{\url{jiri.kvita@upol.cz}}
%


\maketitle


\begin{abstract}

We study the top quark pair events production in $pp$ collisions in the $\ell$+jets channel at the energy of $\sqrt{s} = 14$ TeV for Standard Model as well as new physics processes. We explore the usage of semi-boosted topologies where the top quark decays into a high-transverse momentum (boosted) hadronic $W$-jet and an isolated $b$-jet and study their performance in the $t\bar{t}$ events kinematic reconstruction. An important event fraction is recovered and the correlation of selected kinematic variables between the detector and particle level is studied. Quality of the reconstructed mass line shape of a hypothetical scalar resonance decaying into $t\bar{t}$ is evaluated and compared for regimes of a different degree of the transverse boost. Unfolding performance is checked in terms of comparing the excess of events in spectra before and after the unfolding, concluding with the proof of a signal significance loss after the unfolding procedure for both energy and angle related observables, with possible applications in current LHC experiments.

\end{abstract}
\newpage
\newpage
\section{Introduction}
\label{section:Intro}

This work studies the top quarks pair kinematic reconstruction in a collider detector close to that of the ATLAS detector~\cite{ATLAS} at the Large Hadron Collider (LHC)~\cite{Evans:1129806} at CERN using a parameterized detector simulation provided by \Delphes{}~\cite{deFavereau:2013fsa}. The LHC nominally collides protons at four interaction points where the detectors are located. The two main multi-purpose detector facilities are the ATLAS and CMS~\cite{CMS:2008xjf} detectors which are versatile particle detectors with ability to discern all kinds of particles with the exception of neutrinos. Both are of similar phase-space coverage, resolution and detection and identification capabilities based on different experimental technologies.

Quarks and gluons originating in collisions are not detected directly because they become confined in hadrons or, in case of the top quark, decay before their arrival to the detector. The process of hadronization forms showers of particles collimated in the direction of the original particle, resulting in a hadronic jet reaching the detector. The degree of collimation is proportional to the momentum of the parent particle and this results in particular perpetual positions of hadronic showers in the detector. If the primarily particle has a large momentum with respect to the beam (transverse momentum, $\pt$), the corresponding particle shower is more collimated leading to a reconstructed hadronic final state with an imprint of the parent particle four-vector, including its mass. Also, stable particles of different origin can overlap in a~jet.

The energy of particles used in colliders increases with the advance of the experimental technology. This leads to enrichment of events with particles of higher transverse momenta. This paper studies the process of top and anti-top quark pair($t\bar{t}$) production $pp \rightarrow t\bar{t}$ at the LHC at CERN at the center-of-mass energy $ \sqrt{s} = 14$ TeV. This paper also considers a process with a hypothetical massive heavy scalar particle $y_{0}$ as a mediator for the process $pp \rightarrow y_{0} \rightarrow t\bar{t}$ through a triangle loop for the enhancement of events in the phase space of large transverse momenta.

\section{The $\ttbar$ final states topologies}
\label{section:ttbar}

The $t\bar{t}$ events are categorized into three channels, according to the decaying products of the top quarks. The top quark decay is described by the following process: $t \rightarrow W^{+} q$ $(q = b,s,d)$. The rest of allowed decay processes are weak neutral currents which are heavily suppressed and their contribution is negligible. Furthermore, the decay of the top quark is mainly to the bottom quark thanks to the large value of the CKM mixing matrix element between the bottom and top quarks. The $W$ boson has two main decay modes; hadronic (68\%) and leptonic (32\%)~\cite{PDG}. There are two $W$ boson decays in each $t\bar{t}$ event and the $t\bar{t}$ decay channels can  thus be categorized based on the combination of $W$ decay modes to all-hadronic, semi-leptonic and dilepton channels. This analysis focuses on the semi-leptonic channel.   

\begin{figure}
\begin{center}
\begin{tabular}{cc}

{\includegraphics[width=0.46\textwidth]{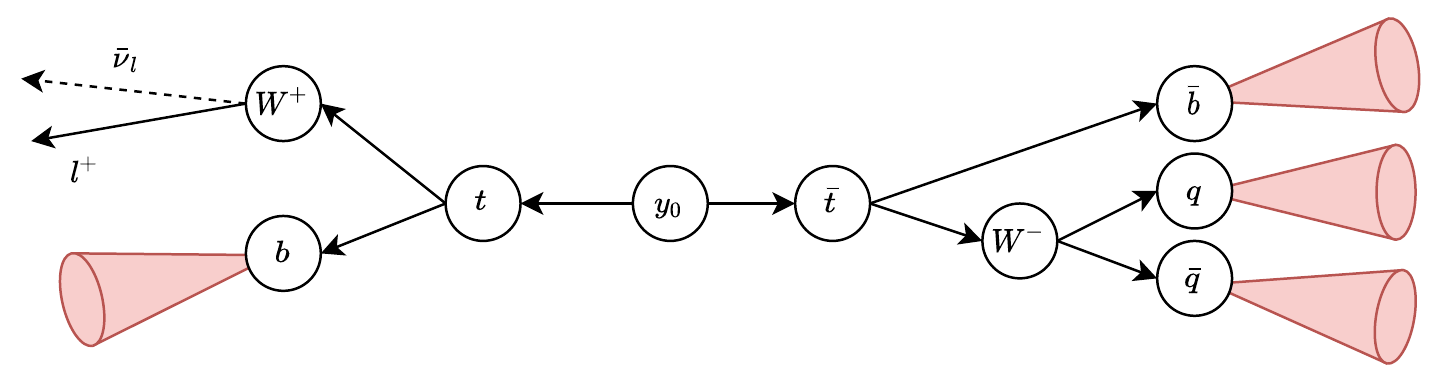}} & \includegraphics[width=0.46\textwidth]{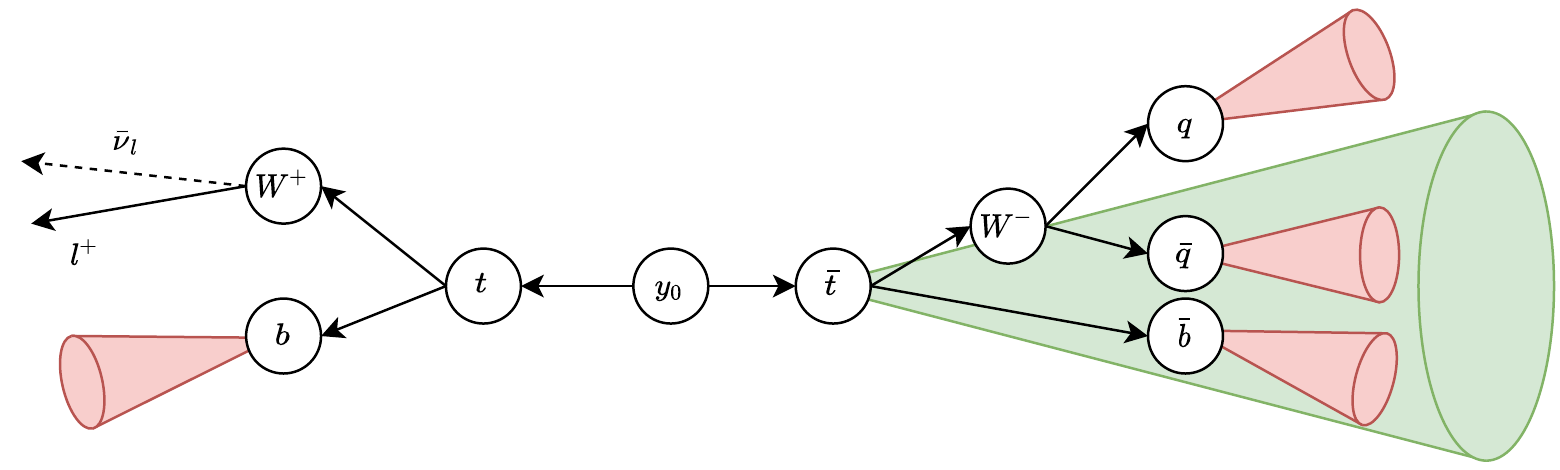}\\
a) & b) \\
\includegraphics[width=0.46\textwidth]{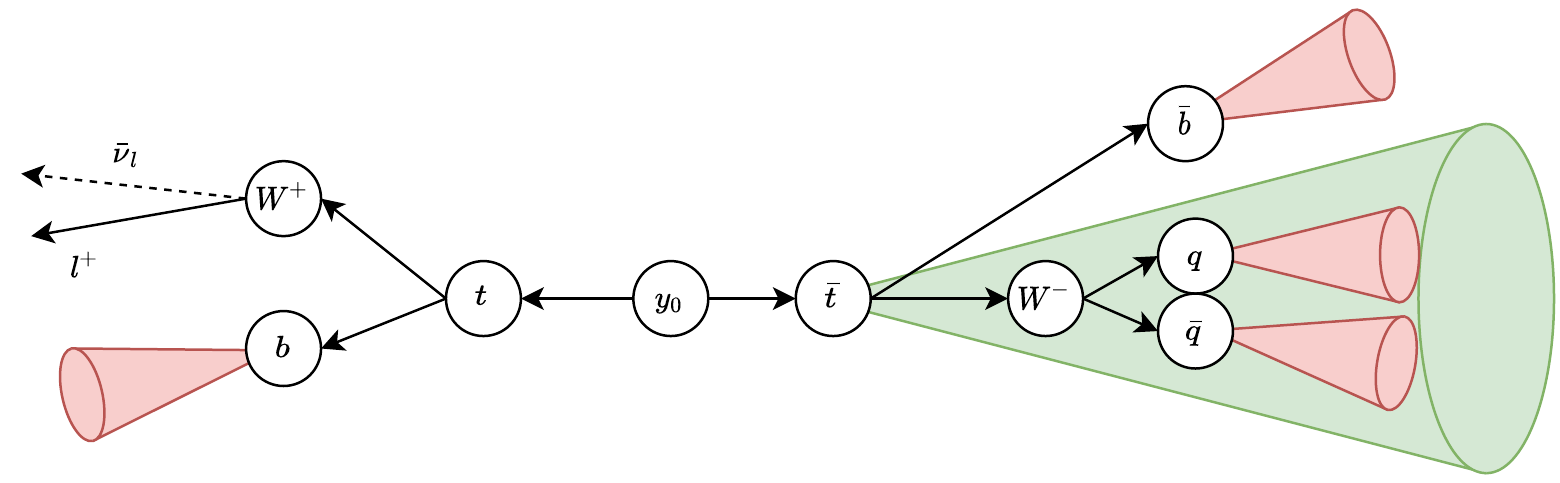} & \includegraphics[width=0.46\textwidth]{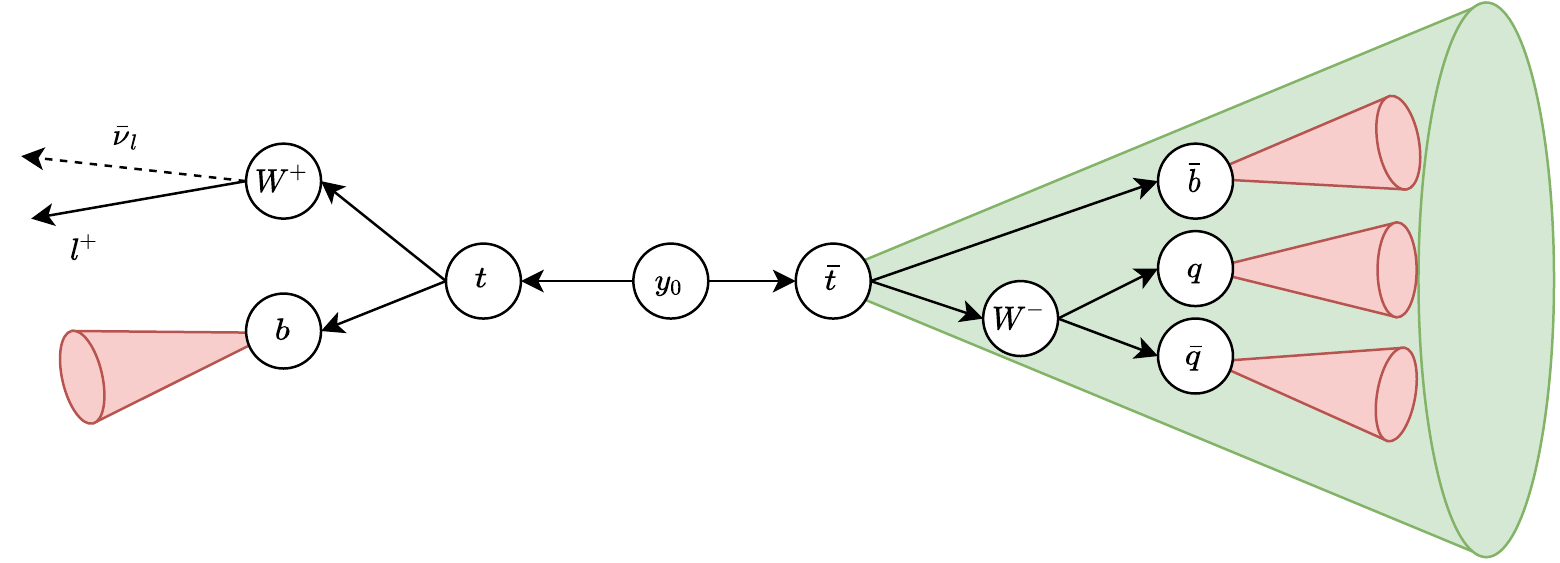}\\
c) & d) \\
\end{tabular}
\caption{A schematic of $pp \rightarrow y_{0} \rightarrow t\bar{t}$ decays in the resolved a), the semi-boosted mixed b), the semi-boosted c) and the boosted d) topologies in the $\ell$+jets channel. Red (green) cones represent small-$R$ (large-$R$) jets.}
\label{fig resolved_top}
\end{center}
\end{figure}

The degree of collimation of the produced particle showers and their angular separation in the detector defines the topology of an event. In the resolved topology, $t\bar{t}$ decay products are reconstructed as individual jets and a lepton, see Fig.~\ref{fig resolved_top}a). Events in this topology are usually produced at lower invariant masses of the $t\bar{t}$ pair. In the semi-boosted topology, decay products on the side where the $W$ boson is decaying hadronically are collimated enough to form one jet in the detector with exception of the jet from the $b$ quark, see Fig.~\ref{fig resolved_top}b). In the semi-boosted mixed topology, which is a special case of semi-boosted topology, the angularly isolated jet is one of the $W$ boson hadronic decay products, see Fig.~\ref{fig resolved_top}c). In the boosted topology, all products from the hadronically decaying top quark are collimated and form one large jet in the detector, see Fig.~\ref{fig resolved_top}d). The fractions of the topologies are correlated to the energy spectrum of the $t\bar{t}$ pair forming a gradual transition from the resolved to boosted topologies. The number of events of the resolved topology drops significantly with increasing energy of the process. In an intermediate energy regime, the number of events in the boosted topology is not yet large enough to fill the gad after the resolved topology.

Finding ways to improve the events reconstruction efficiency by adding the semi-boosted topologies, which reside in the aforementioned transition energy region, is one of the aims of this paper. This can help gain statistics in $t\bar{t}$ analyses. All these four topologies mentioned are explored in this paper. In addition, we study the resolution of the mass peak of an hypothetical scalar resonance decaying to $\ttbar$ and we check the performance of the unfolding procedure in terms of its ability to retain an excess of a possible new physics signal.

The semiboosted topologies have been used in analyses at the LHC as \emph{e.g.} by CMS~\cite{CMS:2014rsx,CMS:2012bti,CMS:2012jea} or ATLAS~\cite{ATLAS:2015lxh,ATLAS:2019kwg,ATLAS:2015irp,ATLAS:2018wis}, although mostly at lower energies or in cases where the boosted $W$-tagged hadronic jet plays a key r\^{o}le in the analysis.
We argue that their potential is still worth exploring in the $\ttbar$ final states also at the highest energies of the LHC to come. We employ the semiboosted topologies in the $\ttbar \rightarrow \ell$+jets final states and emphasize their ability to recover a non-negligible event fraction as well as explore their usage in unfolding differential distributions, \textit{i.e.} in precision measurements as well in searches for new physics.

\section{Samples}
Events were generated for the processes $pp \rightarrow \ttbar{}$ (SM) and $pp \rightarrow y_{0}' \rightarrow \ttbar{}$ with the addition of a~$y_{0}$ scalar particle to the Standard Model \cite{Afik:2018rxl,Arina:2017sng,Albert:2017onk,Kraml:2017atm, Das:2016pbk,Neubert:2015fka,Backovic:2015soa,Mattelaer:2015haa} using the \MadGraph{} version {\tt 2.6.4} simulation toolkit~\cite{Alwall:2014hca}. The spin-0 model \cite{Backovic:2015soa} contains an s-channel color-singlet scalar mediator with purely flavour-diagonal couplings proportional to the masses of the SM particles, therefore leaving top quark as the only relevant SM fermion coupling to the new hypothetical scalar. The parton shower and the hadronization processes were simulated using \Pythia{}8 \cite{Sj_strand_2015}. Masses of the hypothetical $y_{0}$ particle, which serves effectively as a source of semi-boosted and boosted top quarks, were selected as \GeV{500, 600, 700, 800, 900 and 1000} to sample through the region where the number of events in the resolved topology starts to decline rapidly (500~GeV) to where the number of the boosted topology events is becoming dominant (1000~GeV). The decay width of $y_{0}$ of 1\%, 10\% and 30\% of its mass for each sample were studied, results and values in the tables and plots are shown for the decay width of 10\%. 

The SM $t\bar{t}$-sample without the hypothetical $y_{0}$ particle ensures the correspondence with data measured in LHC experiments and is used as a background to $y_0$ and for corrections for the unfolding procedure. The top quark mass in simulation was set to 173 GeV (\MadGraph{} default).

To ensure the strength of the evidence from the contribution of different samples, all samples were weighted to the same luminosity ($\sim 12~\mathrm{fb}^{-1}$) for stacking and unfolding purposes, and which corresponds to the luminosity of the $t\bar{t}$ sample.

The cross-sections of the samples used are summarized in Table~\ref{table_xsec}. The numbers of events in the table are presented for one statistically independent sample and the generated samples also include charge conjugated processes in the decay, \emph{i.e.} the top and anti-top quark decays were swapped.
All samples were generated at the next-to-leading order (NLO) accuracy in perturbative quantum chromodynamics (QCD), allowing also hard process with the additional high-$\pt$ jet production.

The samples for the description of the $W+$jets and $WW+$jets backgrounds were prepared within the same framework as the signal samples. The addition of the associated production of one $W$ boson and two $b$ quarks ($Wbb \rightarrow \ell \nu bb$ + jets) background and two $W$ bosons and two $b$ quarks ($WWbb \rightarrow \ell \nu bb+$jets) background brings the analysis close to those over data while the $y_0$ sample represents a signal of new physics.

The ATLAS-like detector was simulated using the \Delphes{} version {\tt 3.4.1} package \cite{deFavereau:2013fsa} with a modified ATLAS card\footnote{The modification is the addition of information about $B$-hadrons and in the reconstruction of both small as well as large-$R$ jets.}. This simulation is able to perform particle propagation through the magnetic field as well as hadronic and electromagnetic calorimeter simulation including the response of the detector, muon identification system and missing energy. The \Delphes package has its own reconstruction procedure leading to detector-level jets with a simulated realistic energy response based on the performance of the ATLAS detector at LHC.  Jets with two distance parameters 0.4 and 1.0 were reconstructed using the anti-$k_t$ algorithm \cite{antikt2008} to form small-$R$ jets (small jets) and large-$R$ jets (large jets), using the FastJet algorithm~\cite{Cacciari:2011ma}. The \Delphes{} package has its built-in jet energy scale correction for jets, which is in our case used for small jets only as pre-correction, with a private jet energy scale correction applied on top of it for small-$R$ and as a fully private correction for the large-$R$ jets. More details can be found in Appendix~\ref{appendix_JES} and in the selection section. For the large-$R$ jets see Section~\ref{selection:large_jets} while for small-$R$ jets see Section~\ref{section:subsection_small_jet_selection}.

\begin{table}
The cross-section, processes details and the generated number of events for samples generated by the \MadGraph{} package; c.c. stands for the charge conjugation and $\ell$ for an electron or muon. A cut of $p_{\mathrm{T,SJ}}> 20$ GeV is applied at the generator level. In the left column, values on the $y_0$ lines indicate its generated mass.
\begin{center}
\begin{tabular}{@{}lclr@{}} 
\hline\\[-6.5pt]
Sample & Cross-section [pb] & Generated process  & Events\\
\\[-7pt]
\hline
\\[-6pt]
{$Wbb$+jets }& 153.4\hphantom{0000} &{$pp \rightarrow W^{+} + j, W^{+} \rightarrow \ell^{+}+\nu _{\ell}$+c.c.} & 655,855 \\
{$WWbb$+jets }& 180.3\hphantom{0000} &{$pp \rightarrow W^{+} W^{-} b \bar{b} , W^{+} \rightarrow \ell^{+}+\nu _{\ell}, W^{-} \rightarrow jj$+c.c.} & 1,000,000 \\
{$y_{0}$ 1000 GeV} &  0.031 &{$pp \rightarrow y_0 \rightarrow t\bar{t},  t \rightarrow b j j, \bar{t} \rightarrow \bar{b} \ell^{-} \bar{\nu}_{\ell}$+c.c.}   & 500,000 \\
{$y_{0}$ 900 GeV} &  0.053 & {$pp \rightarrow y_0 \rightarrow t\bar{t},  t \rightarrow b j j, \bar{t} \rightarrow \bar{b} \ell^{-} \bar{\nu}_{\ell}$+c.c.}   & 500,000 \\
{$y_{0}$ 800 GeV} &  0.091   & {$pp \rightarrow y_0 \rightarrow t\bar{t},  t \rightarrow b j j, \bar{t} \rightarrow \bar{b} \ell^{-} \bar{\nu}_{\ell}$+c.c.} & 500,000 \\
{$y_{0}$ 700 GeV} &  0.16\hphantom{0}   &{$pp \rightarrow y_0 \rightarrow t\bar{t},  t \rightarrow b j j, \bar{t} \rightarrow \bar{b} \ell^{-} \bar{\nu}_{\ell}$+c.c.}   & 500,000 \\
{$y_{0}$ 600 GeV} &  0.27\hphantom{0}   & {$pp \rightarrow y_0 \rightarrow t\bar{t},  t \rightarrow b j j, \bar{t} \rightarrow \bar{b} \ell^{-} \bar{\nu}_{\ell}$+c.c.}  & 500,000 \\
{$y_{0}$ 500 GeV} &  0.41\hphantom{0}   &{$pp \rightarrow y_0 \rightarrow t\bar{t},  t \rightarrow b j j, \bar{t} \rightarrow \bar{b} \ell^{-} \bar{\nu}_{\ell}$+c.c.}   & 500,000 \\
{$t\bar{t}+$jets} & 178.6\hphantom{0000}   &{$pp \rightarrow t\bar{t},  t \rightarrow b j j, \bar{t} \rightarrow \bar{b} \ell^{-} \bar{\nu}_{\ell}$+c.c.}& 2,934,961 \\ 
\\[-7pt]
\hline
\end{tabular} 
\label{table_xsec}
\end{center}
\end{table}

\section{Object and event selection}
Events considered in the analysis are reconstructed at two levels; once with the \Delphes{} ATLAS-like detector simulation, forming detector level spectra, and at the particle level. The event selection and the requirements differ slightly for the reconstruction level and for the boosted, semi-boosted, semi-boosted mixed and resolved topologies and are described below. Unless stated otherwise, the same object and event selection applies to the particle-level objects and selections.
 
\subsection{Missing transverse energy requirement}
The missing transverse energy ($E_{\mathrm{T,miss}}$) is a measure of energy imbalance in the plane transverse to the beam and is equal to the value of negative vector sum in the transverse plane of energies of all objects leaving a calorimeter deposit. By definition this should equal to zero thanks to the law of energy conservation, but the energy taken away by the undetected neutrinos is not counted for in the detector and their contribution to $\ell/\mu$ channels via leptonic $\tau$ decays is small. The magnitude of the missing transverse energy is required to be $E_{\mathrm{T,miss}} > 25$ GeV for all topologies as well as for both the detector and the particle levels. This ensures that only events in which neutrinos carry away a considerable amount of energy are chosen for the analysis. This is a standard requirement for the missing energy in most of top quark analyses in channels involving a charged lepton.
\subsection{Lepton selection}
A requirement on the lepton (muon or electron) transverse momentum ensures the selected lepton comes from the hard process and a cut of $p_{\mathrm{T},\ell} > 25$ GeV is used, a typical value in real experiment also due to trigger requirements. Tau leptons are not considered in this analysis as they decay before they enter the detector. In case more leptons fulfilling the $\pt$ requirement, only the electron or muon with the highest transverse momentum is taken into account. This requirement is the same for all topologies.

Charged leptons may radiate low energy photons which are highly collimated. \textit{E. g.} for electrons the separation of such photons and the lepton is below the resolution of the detector and thus the photon energy is included by construction at the detector level. The lepton dressing procedure is performed at the particle level reconstruction to correct for this phenomenon, in which the photon four-vectors, fulfilling the condition of the angular separation threshold $\Delta R_{\mathrm{\gamma,\ell}} = \sqrt{\Delta\eta_{\mathrm{\gamma,\ell}}^{2} + \Delta\phi_{\mathrm{\gamma,\ell}}^{2}} < 0.1$, are added to the lepton four-vector.
\subsection{Large jet selection}
\label{selection:large_jets}
Jets are the experimental signatures of hadronic final states of quarks and gluons, which form particle showers entering the detector. In a typical collider detector, jet constituents are clustered energy deposits in calorimeters, or stable particles at the particle level. The jet four-vector is the result of the reconstruction with the anti-$k_{t}$ algorithm.

We call jets reconstructed with a distance parameter $\Delta R = 1$ as large jets or large-$R$ jets. A private jet energy scale correction is derived on the $t\bar{t}$ sample and applied to the detector level large jet before the selection in order to correct jet energies to the particle level. The magnitude of the jet energy scale correction is about 5\% depending on jet pseudorapidity $\eta$\footnote{The pseudorapidity is defined as a function of the polar angle $\theta$ as $\eta \equiv -\ln\tan\frac{\theta}{2}$.} and $p_{\mathrm{T}}$. In the event selection, the transverse momentum of large jets is required to be $p_{\mathrm{T,LJ}} > 100$~GeV. This condition helps to reduce the number of events with jets not coming from top quark or $W$ decays. Furthermore, all large jets are considered in the pseudorapidity range $|\eta| < 2.5$. This constraint ensures in practice better jet identification as the forward (large $|\eta|$) region is not well instrumented for tracking and has a worse energy resolution. The isolation criterion of jets to be isolated from the lepton ensures that the selected lepton is not contained within the large jet by following the requirement of $\Delta R_{\mathrm{LJ,\ell}} = \sqrt{\Delta\eta_{\mathrm{LJ,\ell}}^{2} + \Delta\phi_{\mathrm{LJ,\ell}}^{2}} > 1$. All these requirements are applied to all three topologies\footnote{There is no large jet in the resolved topology.} and both the detector and the particle levels. Each large jet is then probed for the top quark and $W$ boson tagging (see Appendix~\ref{appendix_tagging} for tagging and mistag efficiencies), first for the hypothesis as coming from the top quark decay, then, in the semi-boosted topology, as coming from the $W$ boson decay and in case none of the tagging was successful, the event is then considered as a candidate for the semi-boosted mixed or the resolved topology. 

\begin{figure}
\begin{center}
\begin{tabular}{cc}
\includegraphics[width=0.49\textwidth]{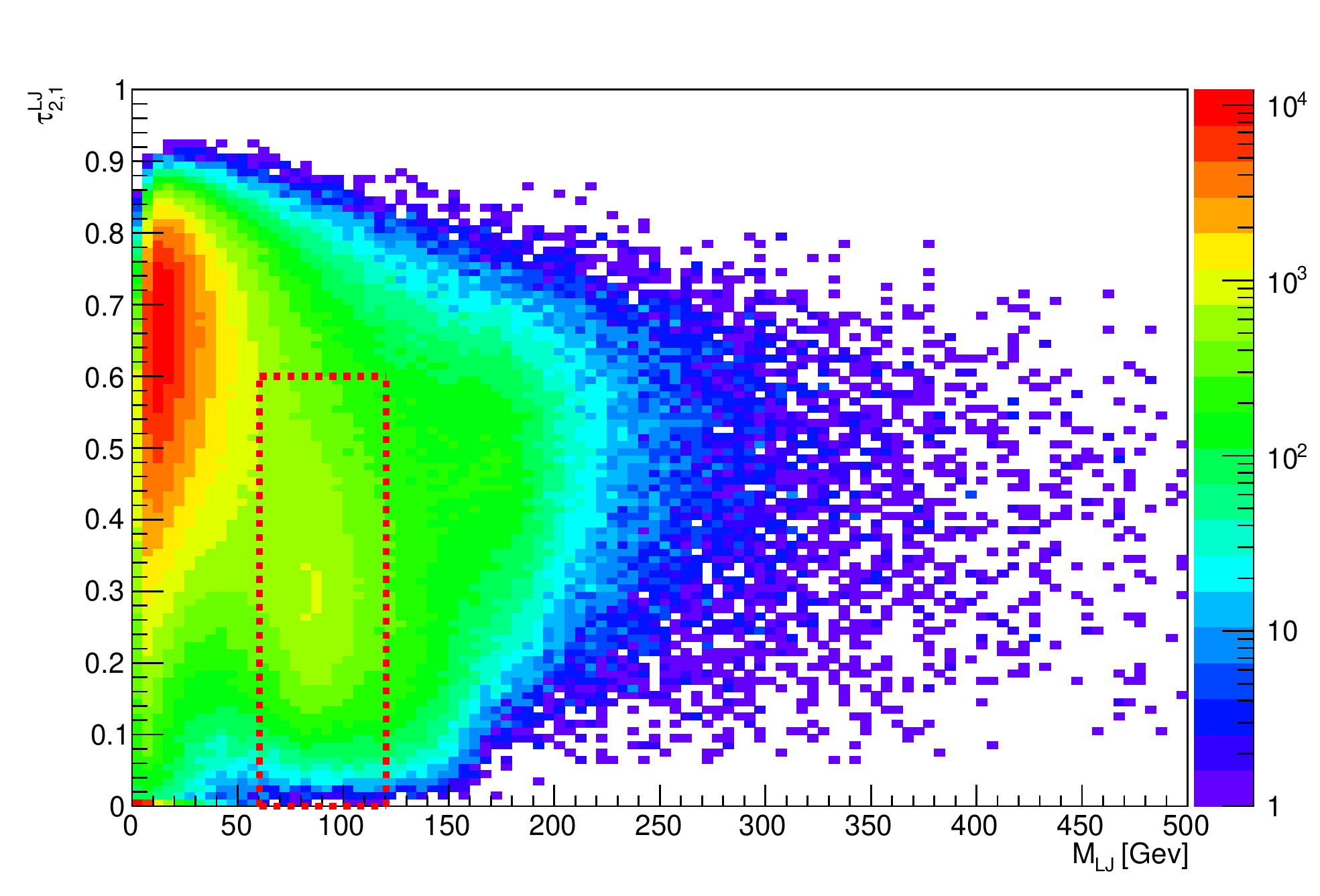}&
\includegraphics[width=0.49\textwidth]{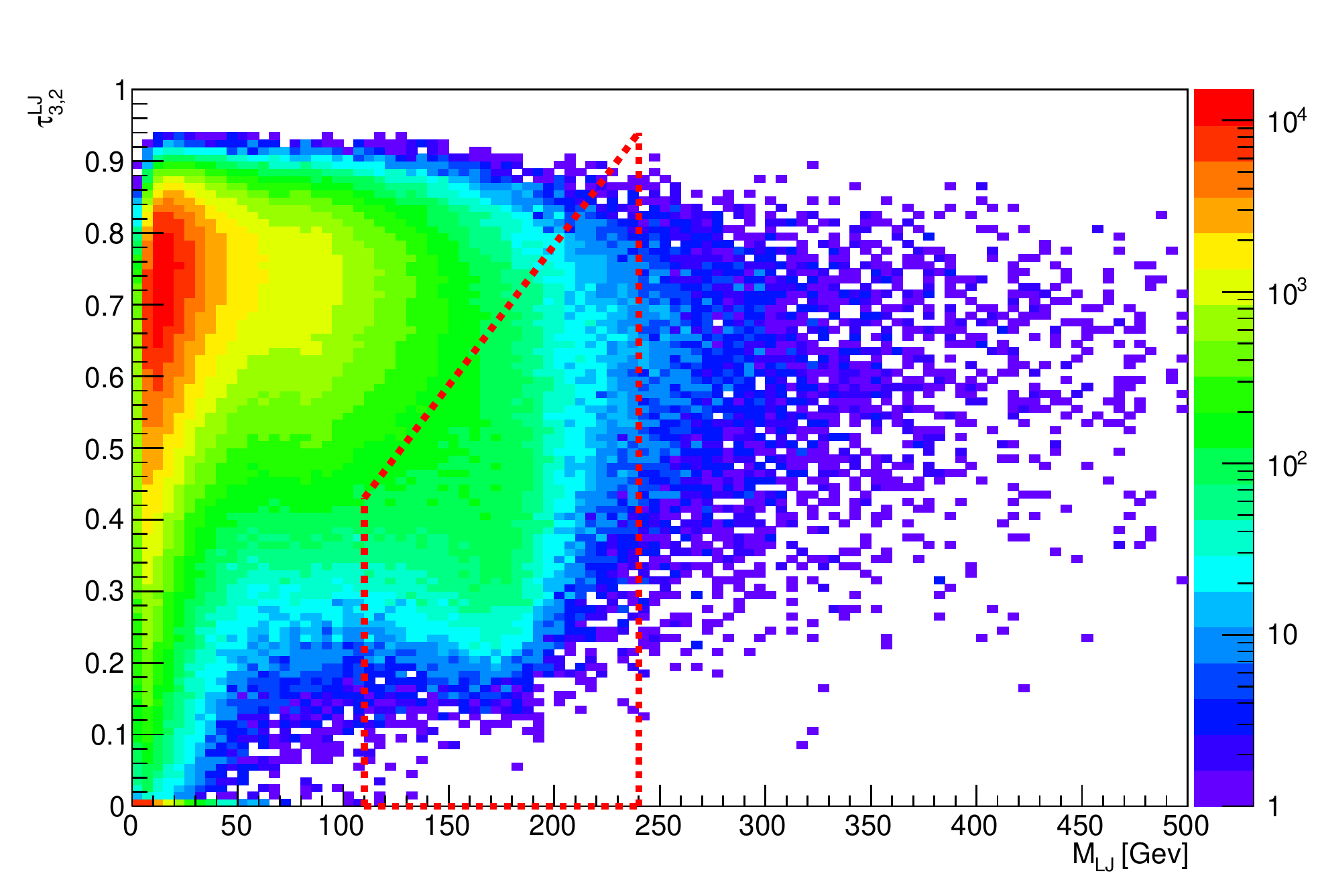}\\

\end{tabular}
\caption{Distribution of the jet substructure variable $\tau_{2,1}$ (left) and $\tau_{3,2}$ (right) in the dependence on the large jet mass ($M_{\mathrm{LJ}}$) for the sample with $M_{\mathrm{y_{0}}}=700$ GeV at the detector level. The large jets in the red dotted area are selected for the reconstruction of the boosted $W$ boson and the top quark. The mass window is set around the expected masses of the $W$ boson and top quark, respectively.} 
\label{fig boosted_tag} 
\end{center}
\end{figure}

Tagging for the boosted topology is based on the constraint on the mass of the large jet \mbox{$110 $ GeV$ < M_{\mathrm{LJ}} < 240$~GeV} and a constraint combining the large jet mass and a jet substructure variable $\tau_{3,2}$ \cite{Nachman:2014kla}, roughly a consistency measure of finding three sub-jets inside the studied large jet rather than two sub-jets, as $M_{\mathrm{LJ}}/\tau_{3,2} > 256$~GeV. The value for the second constraint was added to avoid background events, $\emph{e.g.}$ a large jet from the $W$ boson. The selection is depicted in Fig.~\ref{fig boosted_tag} (right) by the area inside the dotted lines. The jet substructure variable $\tau_{\mathrm{N}}$ is defined as

\begin{equation}
 \tau_{\mathrm{N}}=\frac{1}{d_{\mathrm{0}}}\sum_{k} p_{ \mathrm{T},k}\min{R_{1,k}, R_{2,k}, ...,R_{N,k}},
\end{equation}
\noindent where $d_{0}$ is a normalization parameter computed as
\begin{equation}
 d_{\mathrm{0}} = \Delta R \, \sum_{k} p_{\mathrm{T},k}
\end{equation}
\noindent with $\Delta R$ being the jet distance parameter used (here 1.0). The subjettinesses are then combined into a  ratio, such as $\tau_{2,1}=\frac{\tau_{2}}{\tau_{1}}$, which has the ability to distinguish the compatibility of the jet substructure with the one subjet hypothesis (ratio closer to unity) in comparison to the scenario with two subjets (ratio closer to zero). The substructure variable $\tau_{3,2}$ is defined in a similar manner and effectively compares the jet substructure consistency with two or three subjets.

Tagging of large jets for the semi-boosted topology is based on the large jet mass window $60$~GeV $< M_{\mathrm{LJ}} < 120$~GeV and the jet substructure variable $\tau_{2,1}$, reflecting the consistency of the jet to contain two sub-jets inside the studied large jets rather than a single sub-jet, as $\tau_{2,1} < 0.6$. The selection is shown in Fig.~\ref{fig boosted_tag} (left). 

The large jet in the semi-boosted topology can be tagged as coming from the $W$ boson, but there is no expected peak in the large jet mass spectrum in the semi-boosted mixed topology, and thus it cannot be tagged based on its mass.

The case of the semi-boosted mixed topology is also considered, where the isolated small jet does not originate from the $b$-quark but from a light quark from the $W$ boson decay. To ensure the consistency with a $b$-quark being a part of the large jet, there is a requirement on the large jet to overlap with a small jet originating from a $b$-quark by requiring the condition $\Delta R_{\mathrm{LJ,SJ}} = \sqrt{\Delta\eta_{\mathrm{LJ,SJ}}^{2} + \Delta\phi_{\mathrm{LJ,SJ}}^{2}} < 0.5$ at the detector level, see Section \ref{section:subsection_small_jet_selection} for further information about small jets. A similar condition is set at the particle level for the large jet to contain a $B$-hadron within $\Delta R_{\mathrm{LJ,B-had}} < 0.5$.

\subsection{Small jets selection}
\label{section:subsection_small_jet_selection}

The general requirements for small jets (small-$R$ jets), which are jets reconstructed with a distance parameter $\Delta R = 0.4$, are on the transverse momentum $p_{T} > 25$~GeV, pseudorapidity $|\eta| < 2.5$ and the isolation from the selected lepton $\Delta R_{\mathrm{SJ,\ell}} > 0.5$. The jet energy scale correction is applied on the detector level small jet objects before the selection, which was derived on the $t\bar{t}$ sample on top of the \Delphes{} default jet energy scale. The magnitude of this residual jet energy scale correction is about 2\%.
The identification of the small jets as coming from the $b$-quark, $b$-tagging, is done by the \Delphes{} simulation at the detector level using the efficiency parameterization taken from~\cite{ATL-PHYS-PUB-2015-022}, leading to the $b$-tagging efficiency of 67.4\% (72.8\%) for jets of $\pt=50$~(100)~GeV.
At the particle level, $b$-tagging is done by the requirement of containing a $B$-hadron within the jet as $\Delta R_{\mathrm{SJ,B-had}} < 0.2$ for $B$ hadrons of $\pt > 5$ GeV as recommended by the LHC Top WG \cite{LHCTopWG}.

\subsubsection{Small jet for the reconstruction of the leptonically decaying top quark}
\label{selection:subsection_small_jet_lep}
The small jet for the reconstruction of the leptonically decaying top quark has to fulfill the angular condition $\Delta R_{\mathrm{SJ,\ell}} < 2$, which ensures that it lies in the vicinity of the selected lepton, and must be $b$-tagged. The condition of a large jet isolation from the lepton $\Delta R_{\mathrm{SJ,LJ}} > 1.5$ applies to all topologies with the exception of the resolved topology, where there is no large jet. Such a selected small jet is then removed from the jet collection and from further consideration. This is the only selected small jet in case of the boosted topology.

\subsubsection{Small jet for the hadronically decaying top quark, semi-boosted topology}
\label{selection:subsection_small_jet_had_SB}

For the reconstruction of the hadronically decaying top quark in the semi-boosted topology a small $b$-tagged jet is required in the vicinity to the selected large jet $1 < \Delta R_{\mathrm{SJ,LJ}} < 1.5$. Thus a partial overlap between the selected large jet and the considered small jet is allowed, \emph{i.e.} the selected $b$-tagged small jet should be partially contained in the selected large jet. 

\subsubsection{Small jet for the hadronically decaying top quark, semi-boosted mixed topology}
\label{selection:subsection_small_jet_had_SBM}
The conditions for the semi-boosted mixed topology are similar to the conditions for the semi-boosted topology. The vicinity condition to the large jet remains unchanged but the small jet is required not to be $b$-tagged while a $b$-tag is required for the selected large jet.

\subsubsection{Small jet for the hadronically decaying top quark, resolved topology}
\label{selection:subsection_small_jet_had_R}
The resolved topology selection is tried as the last option before the event is discarded. The reconstruction of the hadronically decaying top quark in the resolved topology requires three small jets, one of them $b$-tagged. The algorithm first takes two small non-$b$-tagged jets with the highest transverse momentum and tests their invariant mass $M_{\mathrm{SJ,SJ}} < 120$ GeV to avoid dijets not corresponding to the mass of the $W$ boson. Then it adds the four-vector of the remaining $b$-tagged jet\footnote{One $b$-tagged jet is used in the reconstruction of the leptonically decaying top quark.}. If all such three jets are found, the event is accepted. 

\section{Reconstruction}

The events passing the selection described in the previous chapter are entering the top anti-top quark pair four-vector reconstruction described in this section and illustrated on the $y_{0} \rightarrow t\bar{t}$ sample with $M_{\mathrm{y_{0}}} = 700$ GeV, although all samples were processed the same way. 

\subsection{Leptonically decaying top quark}

\begin{figure}[!h]
\begin{center}
\begin{tabular}{cc}
\includegraphics[width=0.49\textwidth]{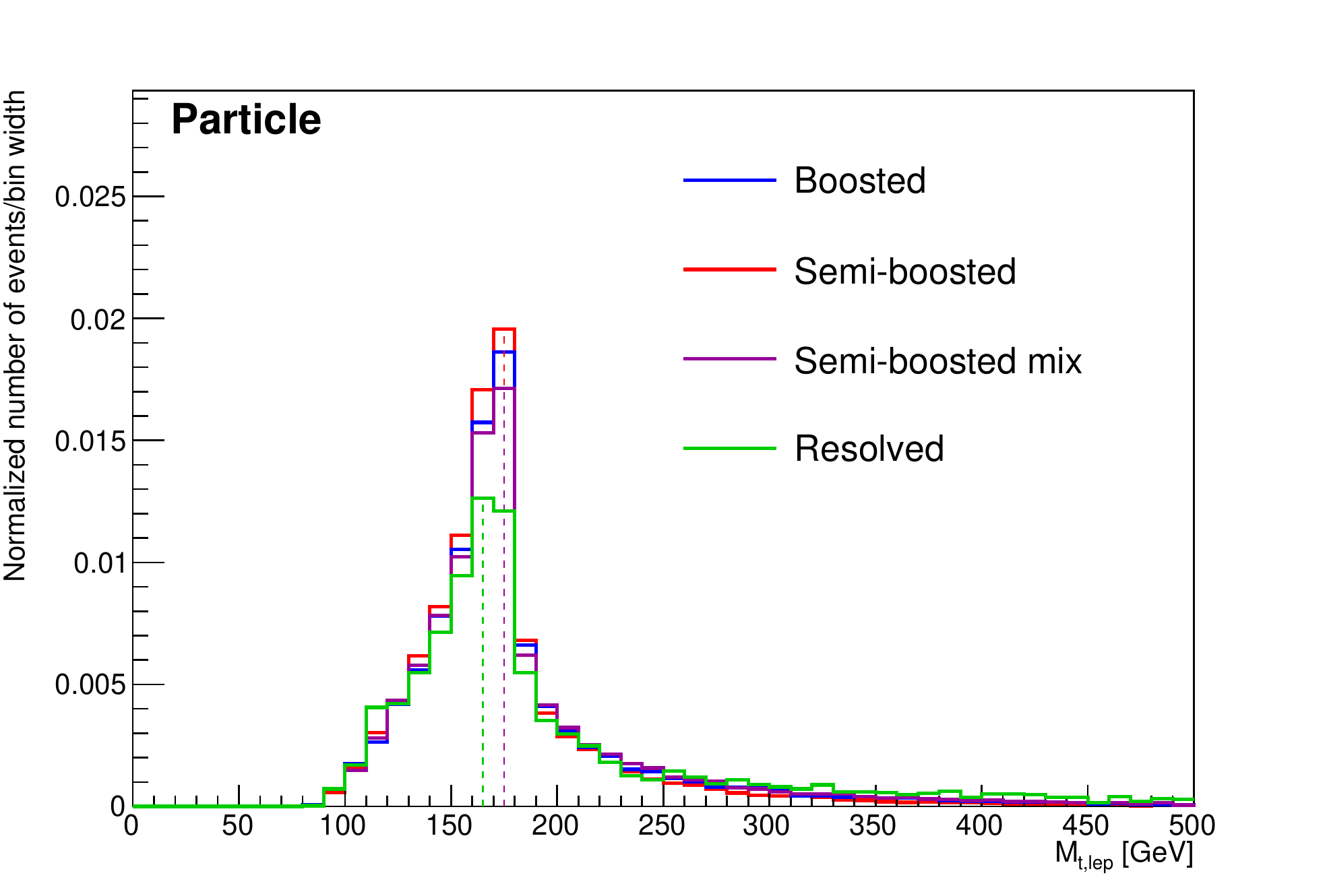} &
\includegraphics[width=0.49\textwidth]{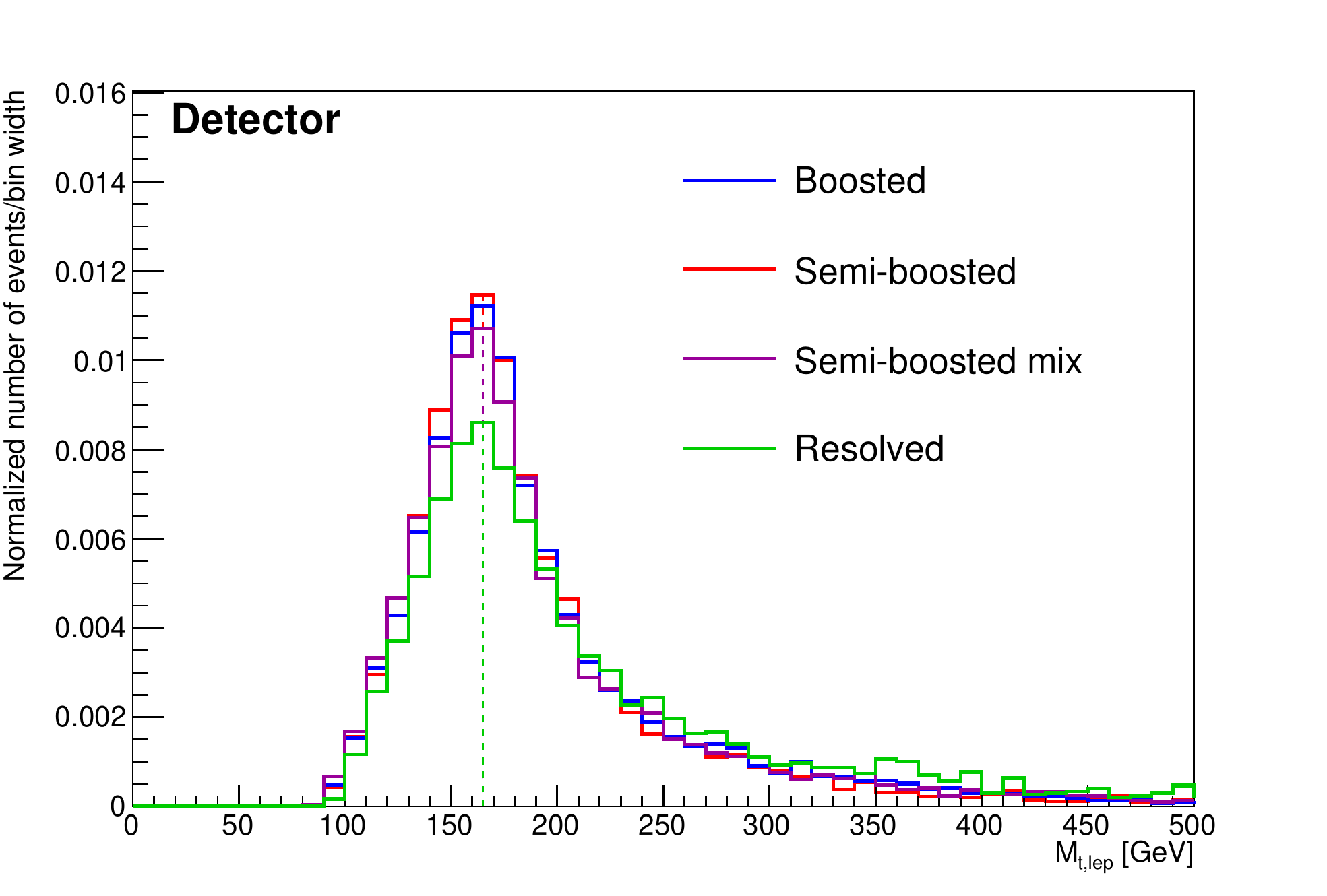} \\
\end{tabular}
\caption{Comparison between topologies for the shapes of the reconstructed leptonically decaying top quark mass ($M_{\mathrm{t,lep}}$) for the sample with $M_{\mathrm{y_{0}}} = 700$ GeV at the particle (left) and the detector (right) levels. The vertical dashed lines indicate the position of the maximum value in the spectrum for each of the topologies, which happens to be the same bin for all topologies at the particle level.}
\label{Wl_rec_lep}
\end{center}
\end{figure}

The reconstruction of the leptonically decaying top quark is the same for all four studied topologies, starting with setting the transverse momentum of the neutrino ($p_{\mathrm{T,\nu}}$) with the missing transverse energy $E_{\mathrm{T,miss}}$. The missing energy together with the four-vector of the selected lepton is used to calculate the longitudinal component of the neutrino momentum  ($p_{z,\nu}$) from the $W$ boson mass constraint  $M_{W} = M_{\ell\nu}$, which leads to a quadratic equation with two solutions in general. The solution which leads to the more central neutrino in the rapidity is accepted in the reconstruction. If the solution leads to a complex number result, the imaginary part is discarded. This procedure is often used in top quark analyses, \textit{e.g.} in the ATLAS experiment \cite{ATLAS}. The $W$ boson is reconstructed as the sum of four-vectors of the lepton and the reconstructed neutrino. Finally, the top quark four-vector is formed from the reconstructed $W$ boson four-vector and the selected $b$-tagged small jet as described in Section~\ref{selection:subsection_small_jet_lep}. The mass of the reconstructed leptonically decaying top quark in the studied topologies is shown in Fig.~\ref{Wl_rec_lep} at both the detector and particles levels.

\subsection{Hadronically decaying top quark, boosted topology}
The recognition of the boosted event is done by selecting a large jet fulfilling conditions specified in Section~\ref{selection:large_jets}. Since there is no reconstructed $W$ boson candidate in this case, the top-tagged large jet is considered to contain most of the products coming from the top quark decay. To verify this, the mass of the large jet corresponding to the reconstructed hadronically decaying top quark is shown in Fig.~\ref{top_rec_res} at both the detector and the particle levels, showing a peak around the top quark mass. 

\subsection{Hadronically decaying top quark, semi-boosted topology}
The reconstruction of the hadronically decaying top quark in the semi-boosted topology uses the selected $W$-tagged large jet and one small $b$-tagged jet, fulfilling the conditions mentioned in Section~\ref{selection:subsection_small_jet_had_SB}. The large jet is considered as the hadronically decaying $W$ boson, with its mass shown in Fig.~\ref{Wh_rec_res}. The reconstructed top quark is formed by adding the selected $b$-tagged small jet four-vector and its mass is shown in Fig.~\ref{top_rec_res}, again exhibiting the expected peak which is sharper at the particle level due to finite detector resolution.

\subsection{Hadronically decaying top quark, semi-boosted mixed topology}
The reconstruction of the hadronically decaying top quark in the semi-boosted mixed topology is performed by summing one large jet and one non-$b$-tagged small jet four-vectors, see Section~\ref{selection:subsection_small_jet_had_SBM} for details. The invariant mass of such a large jet does not produce a peak but in the combination with the selected small jet the resulting invariant mass peak should correspond to the one of the top quark as shown in Fig.~\ref{top_rec_res}.

\subsection{Hadronically decaying top quark, resolved topology}
\begin{figure}
\begin{center}
\begin{tabular}{cc}
\includegraphics[width=0.49\textwidth]{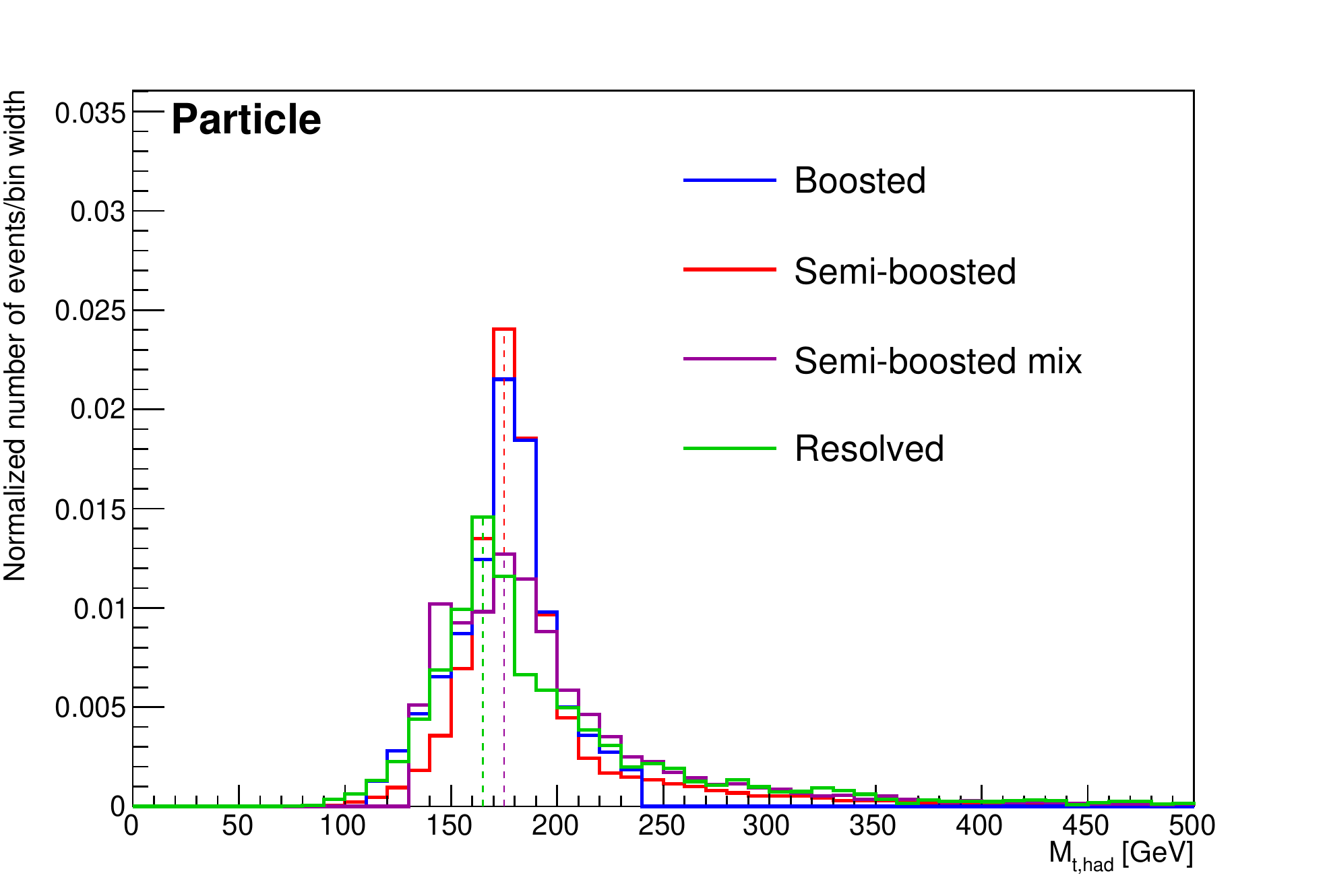}&
\includegraphics[width=0.49\textwidth]{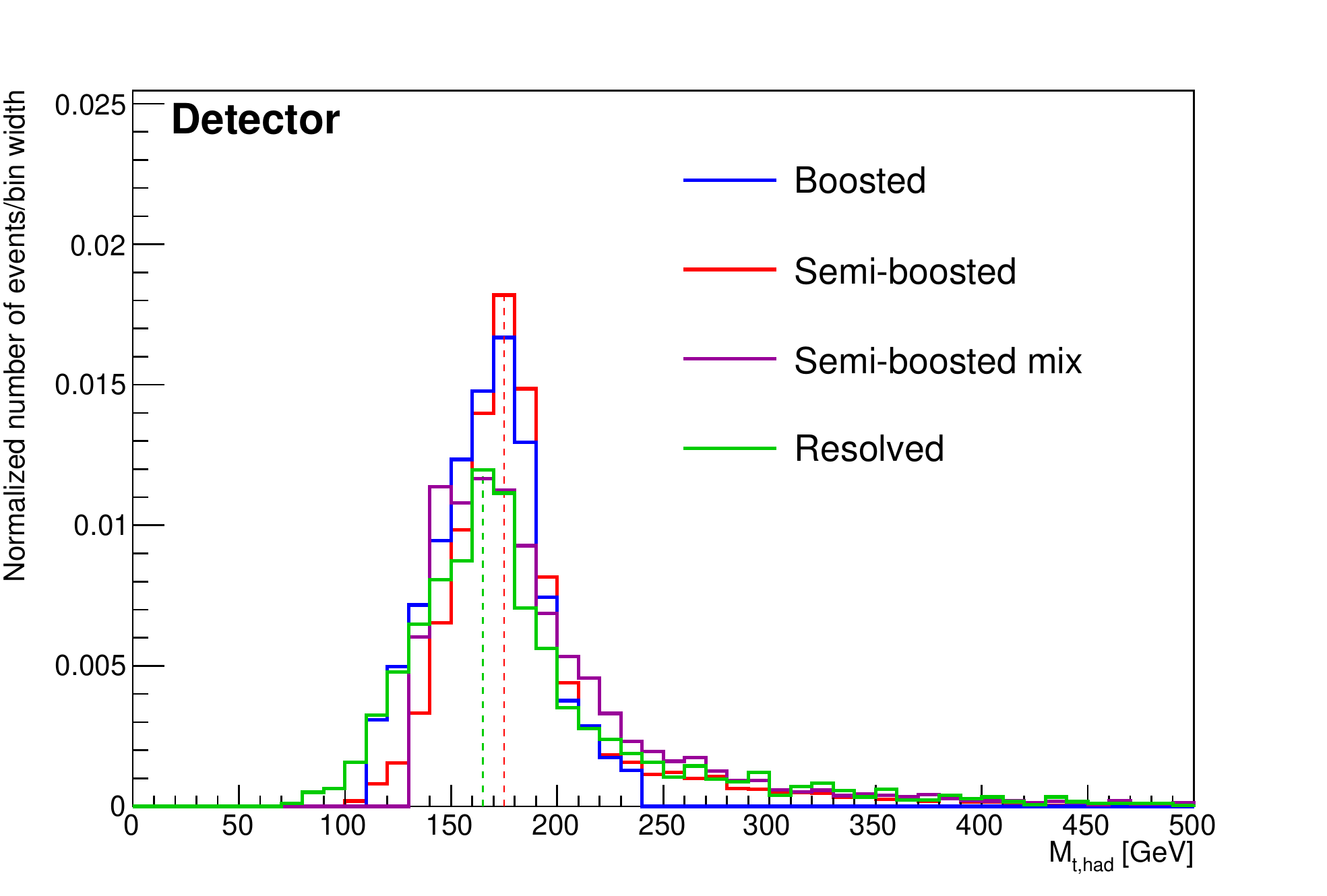} \\
\end{tabular}
\caption{Comparison between topologies for the hadronically decaying top quark mass ($M_{\mathrm{t,had}}$) for the sample with $M_{\mathrm{y_{0}}} = 700$ GeV at the particle (left) and the detector (right) level. The vertical dashed lines indicate the position of the maximum value in the spectrum for each of the topologies.}
\label{top_rec_res}
\end{center}
\end{figure}

\begin{figure}
\begin{center}
\begin{tabular}{cc}
\includegraphics[width=0.49\textwidth]{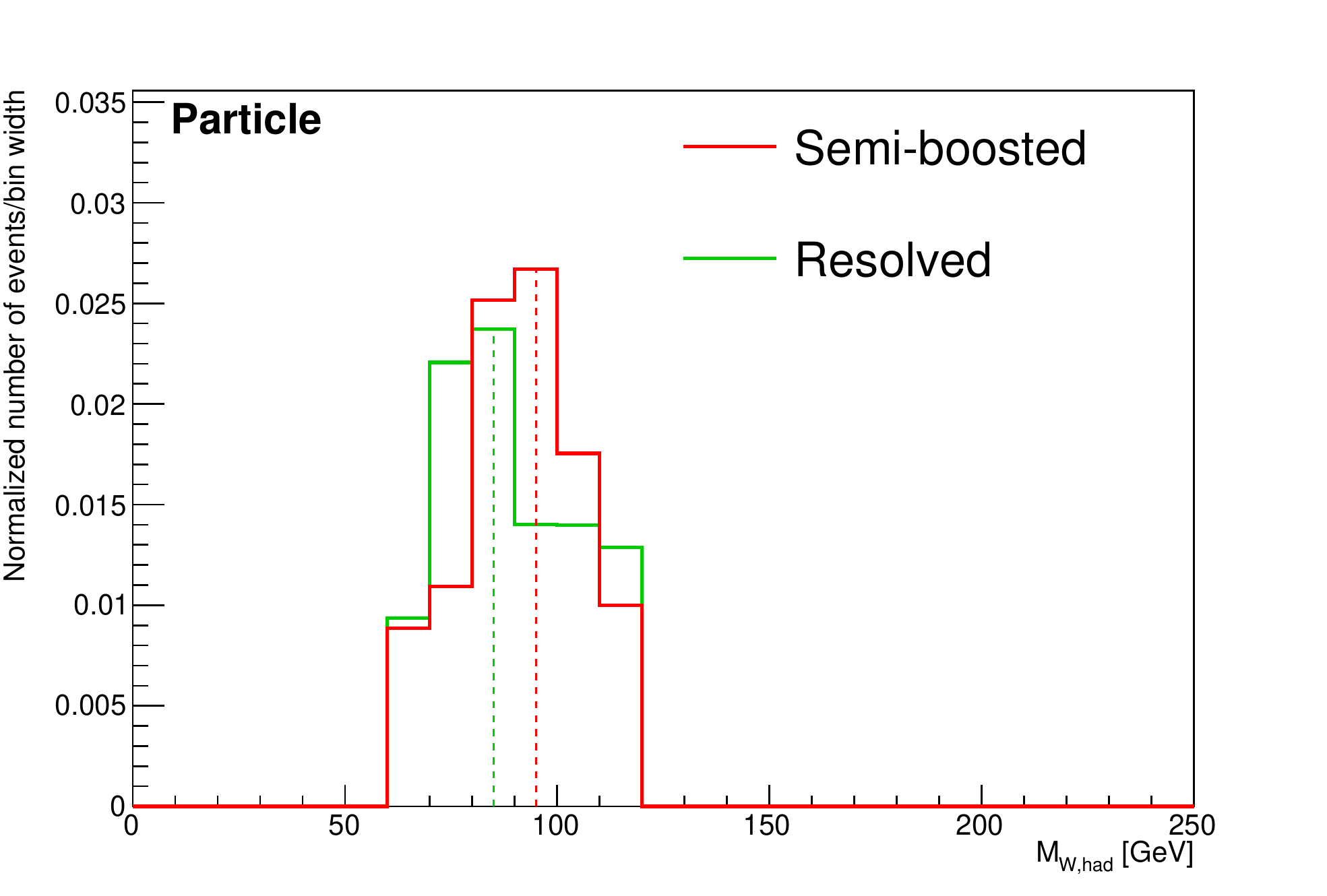}&
\includegraphics[width=0.49\textwidth]{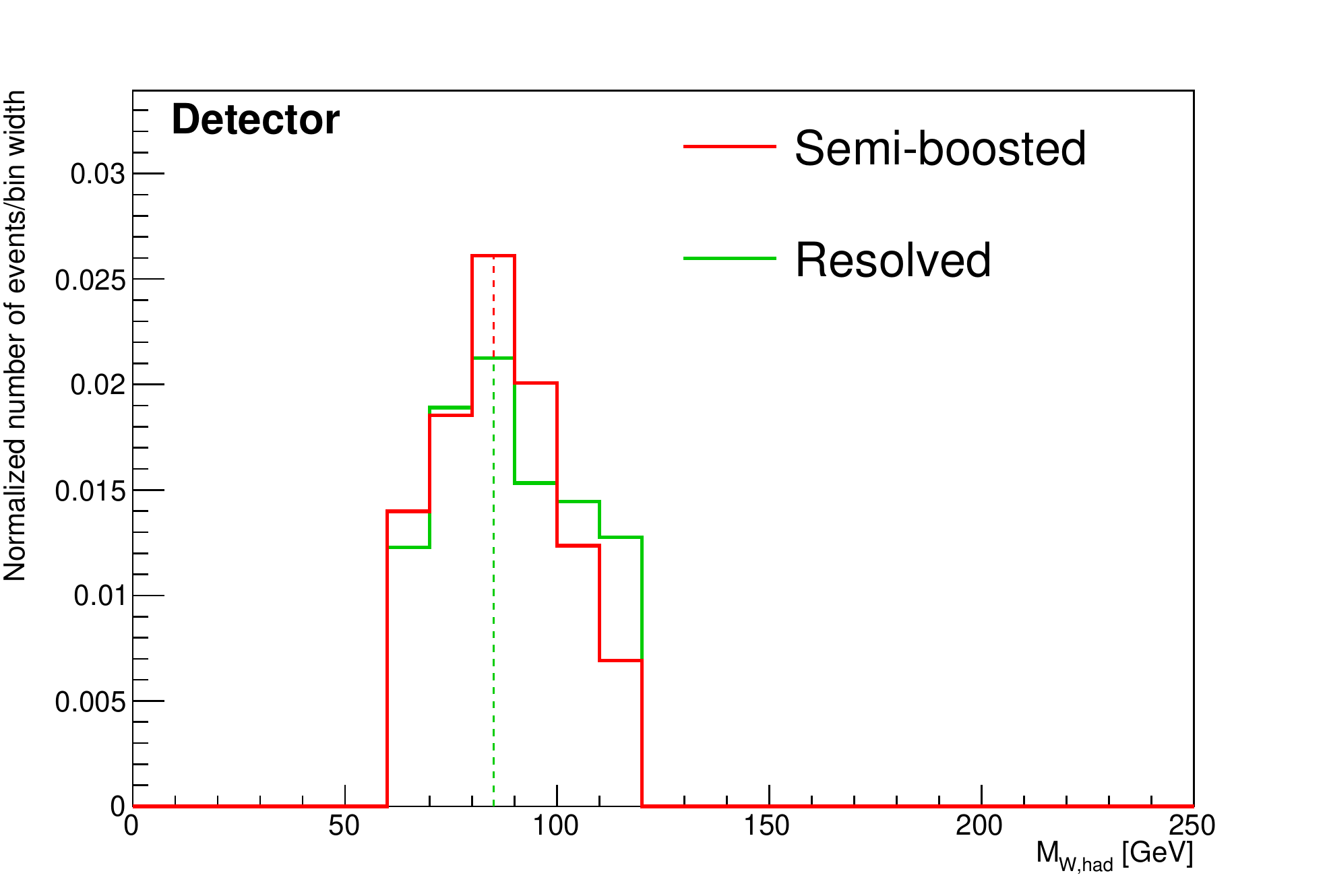} \\
\end{tabular}
\caption{Comparison between topologies for the shapes of the reconstructed hadronically decaying $W$ boson mass ($M_{\mathrm{W,had}}$) for the sample with $M_{\mathrm{y_{0}}} = 700$ GeV at the particle level (left) and the detector level (right). The vertical dashed lines indicate the position of the maximum value in the spectrum for each of the topologies.}
\label{Wh_rec_res}
\end{center}
\end{figure}

The resolved topology has the largest combinatorial ambiguity as it involves largest multiplicity of objects for the reconstruction, starting with the $W$ boson reconstruction from two  highest transverse momentum small jets which are not tagged as $b$-jets. The reconstructed $W$ boson candidate mass, shown in Fig.~\ref{Wh_rec_res}, is required to be 60--120 GeV, otherwise the event is discarded. The third selected small jet for the reconstruction in the resolved topology is required to be $b$-tagged and is added to the reconstructed $W$ boson forming finally the four-vector of the hadronically decaying top quark with its mass shown in Fig.~\ref{top_rec_res}.

\subsection{Top anti-top quark pair system}
\begin{figure}
\begin{center}
\begin{tabular}{cc}
\includegraphics[width=0.49\textwidth]{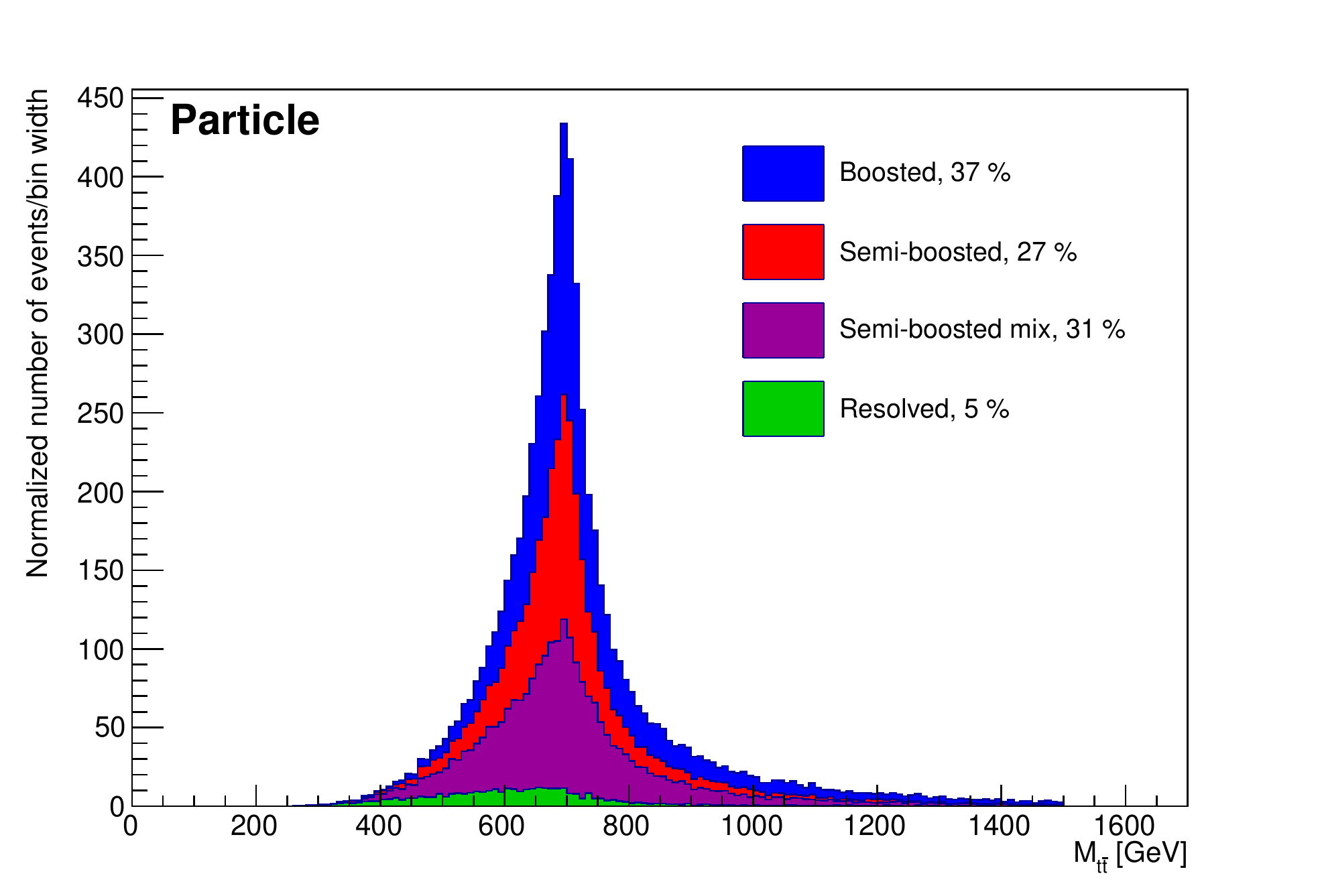} &
\includegraphics[width=0.49\textwidth]{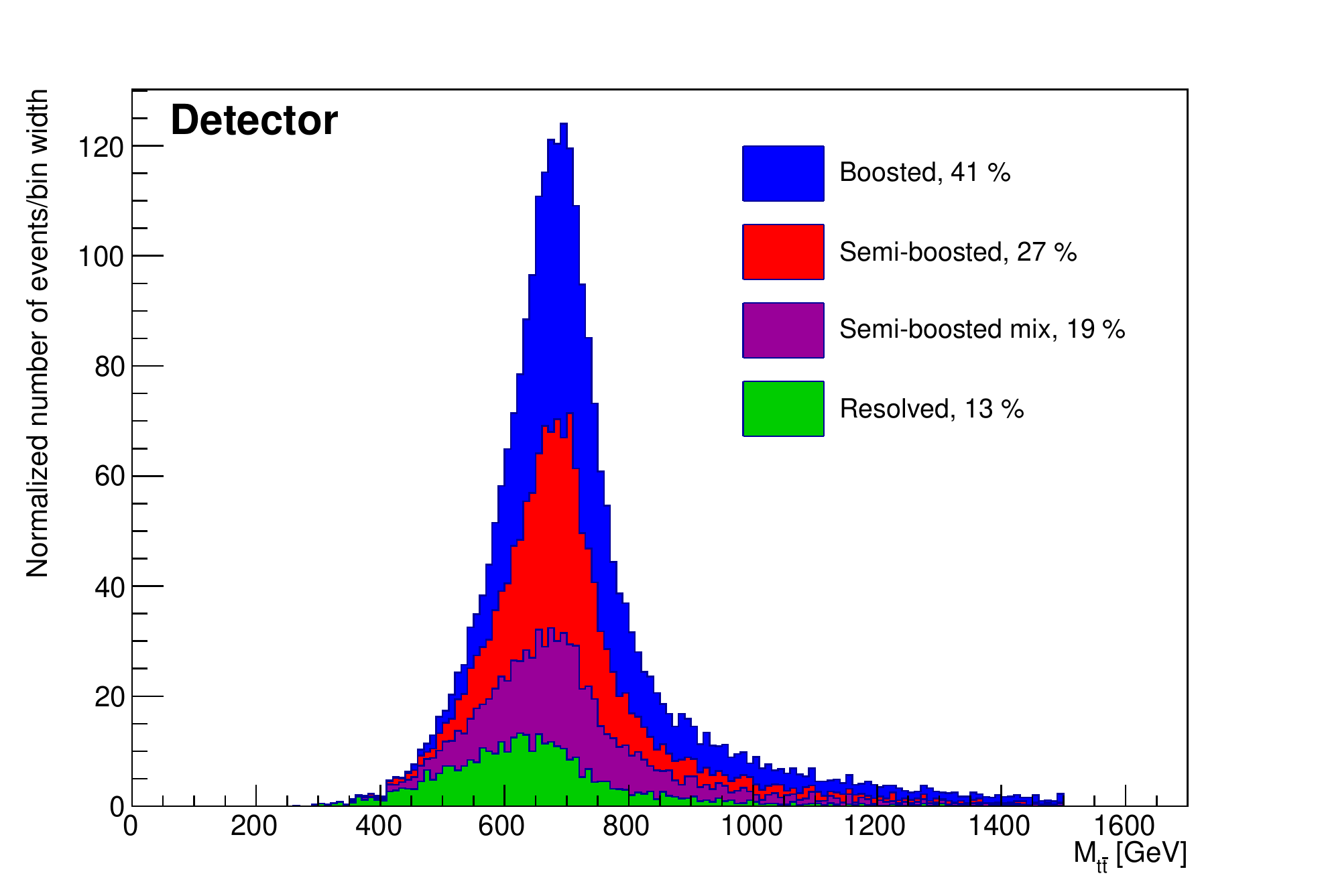} \\
\end{tabular}
\caption{Contributions of different topologies to the reconstruction of the $\ttbar$ invariant mass ($M_{\mathrm{t\bar{t}}}$) for the sample with $M_{\mathrm{y_0}} = 700$ GeV in descendant order: boosted (blue), semi-boosted (red), semi-boosted mixed (purple), and resolved (green) at the particle (left) and the detector (right) level. The corresponding percentage is presented in the legend for each topology.}
\label{fig_rec_sys_stack}
\end{center}
\end{figure}

\begin{figure}
\begin{center}
\begin{tabular}{c}
\includegraphics[width=0.70\textwidth]{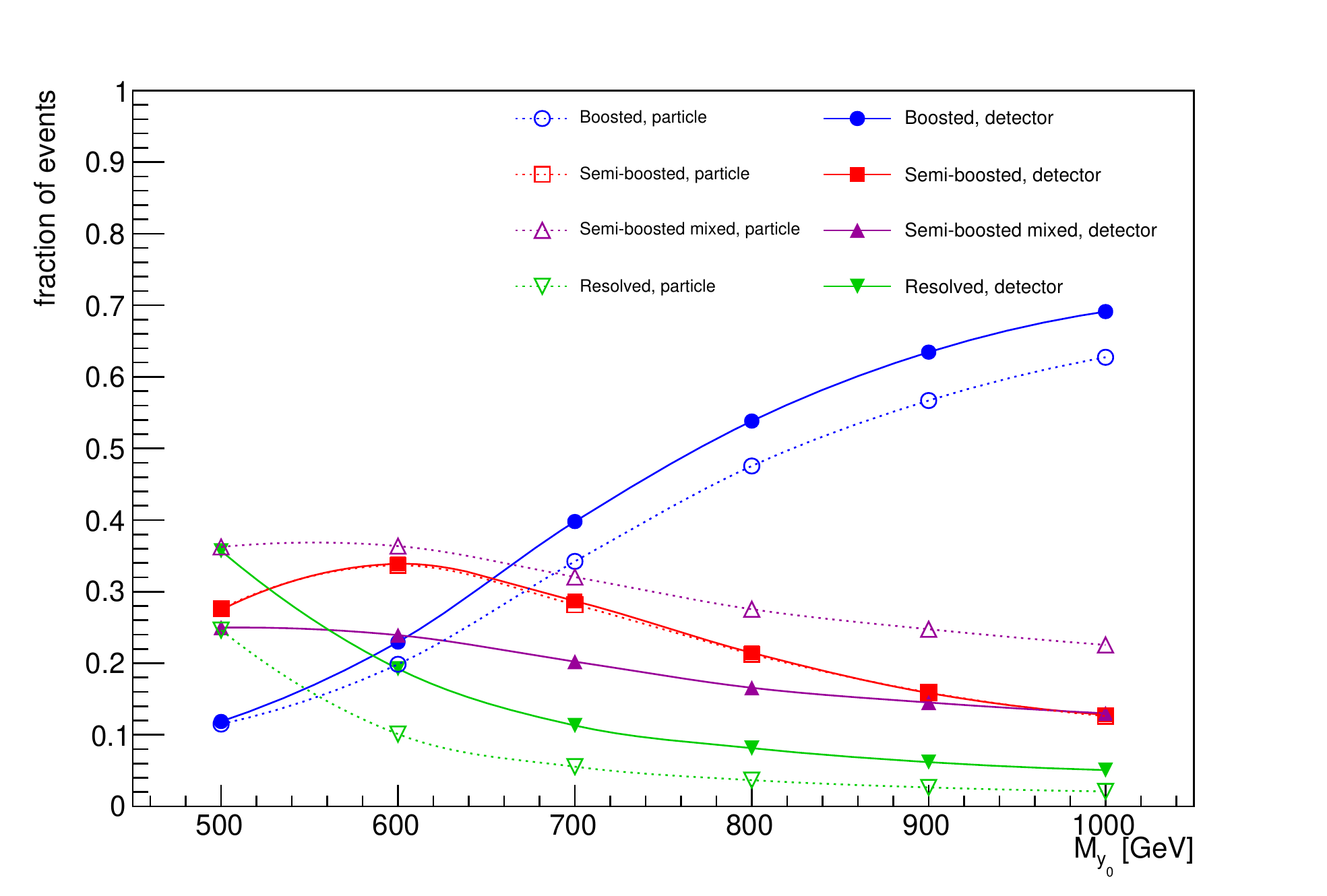} \\
\end{tabular}
\caption{The fraction of events contributing to the $t\bar{t}$ reconstruction from each topology over samples with various masses of the hypothetical $y_{0}$ particle ($M_{y_{0}}$) at the detector (solid lines, full markers) and particle (dotted lines, open markers) levels.}
\label{fig_rec_sys_stack_over_samples} 
\end{center}
\end{figure}

The four-vector of the top anti-top quark ($\ttbar{}$) pair system is reconstructed as the combination of the leptonically and  hadronically decaying top quarks. Its mass and the contributing fractions of events from the particular topologies are shown in Fig.~\ref{fig_rec_sys_stack}. The fractions depend on the mass of the hypothetical $y_0$ particle as shown in Fig.~\ref{fig_rec_sys_stack_over_samples}. This plot illustrates the existence of the transition region between the resolved and the boosted topology which was mentioned in Section~\ref{section:Intro} and which benefits from the implementation of the semi-boosted and semi-boosted mixed topologies via their non-negligible event fractions. 

\subsection{Migration of events between topologies}

\begin{figure}  
\begin{center}
\begin{tabular}{cc}
\includegraphics [width=0.49\textwidth]{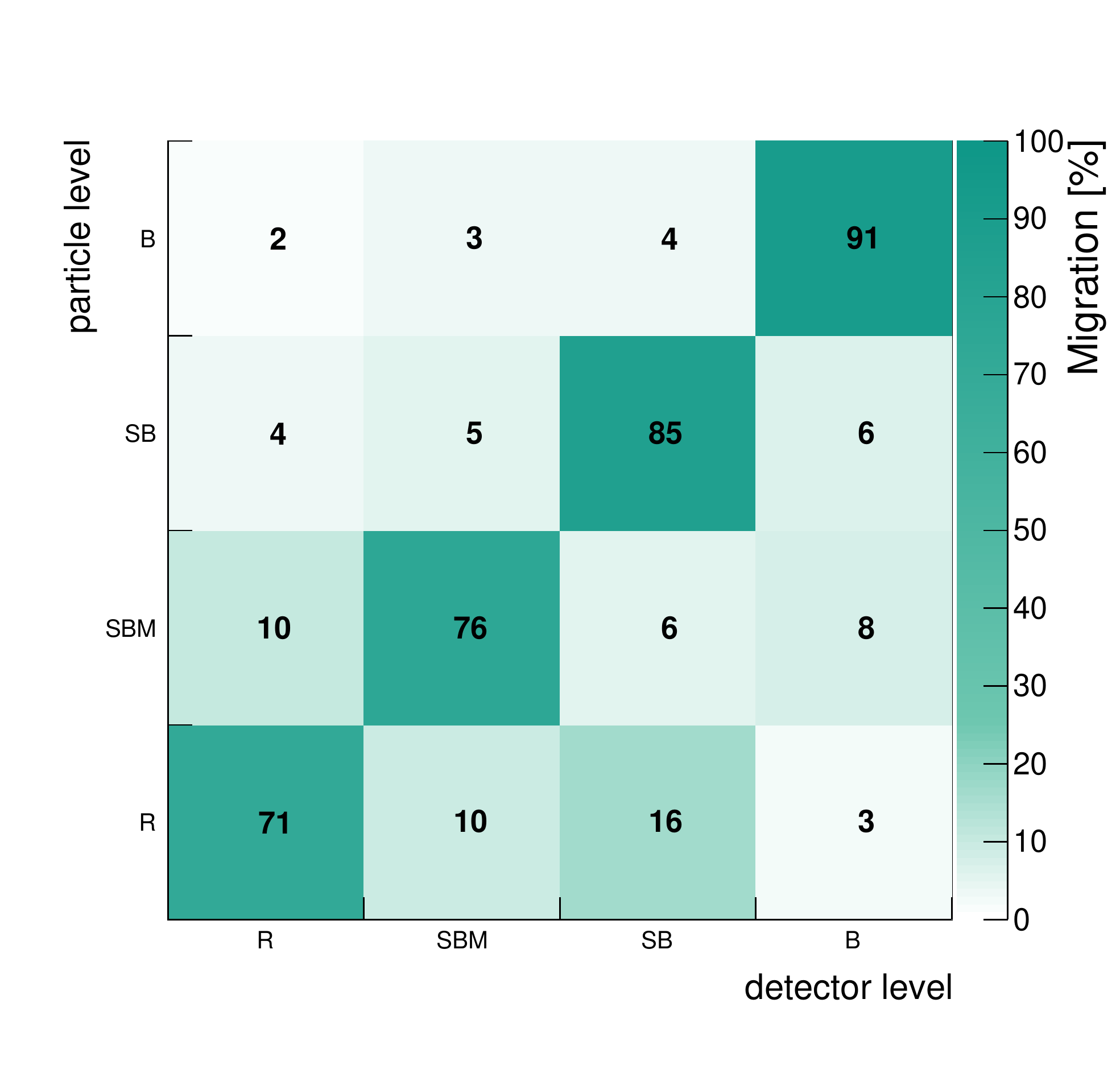} &
\includegraphics [width=0.46\textwidth]{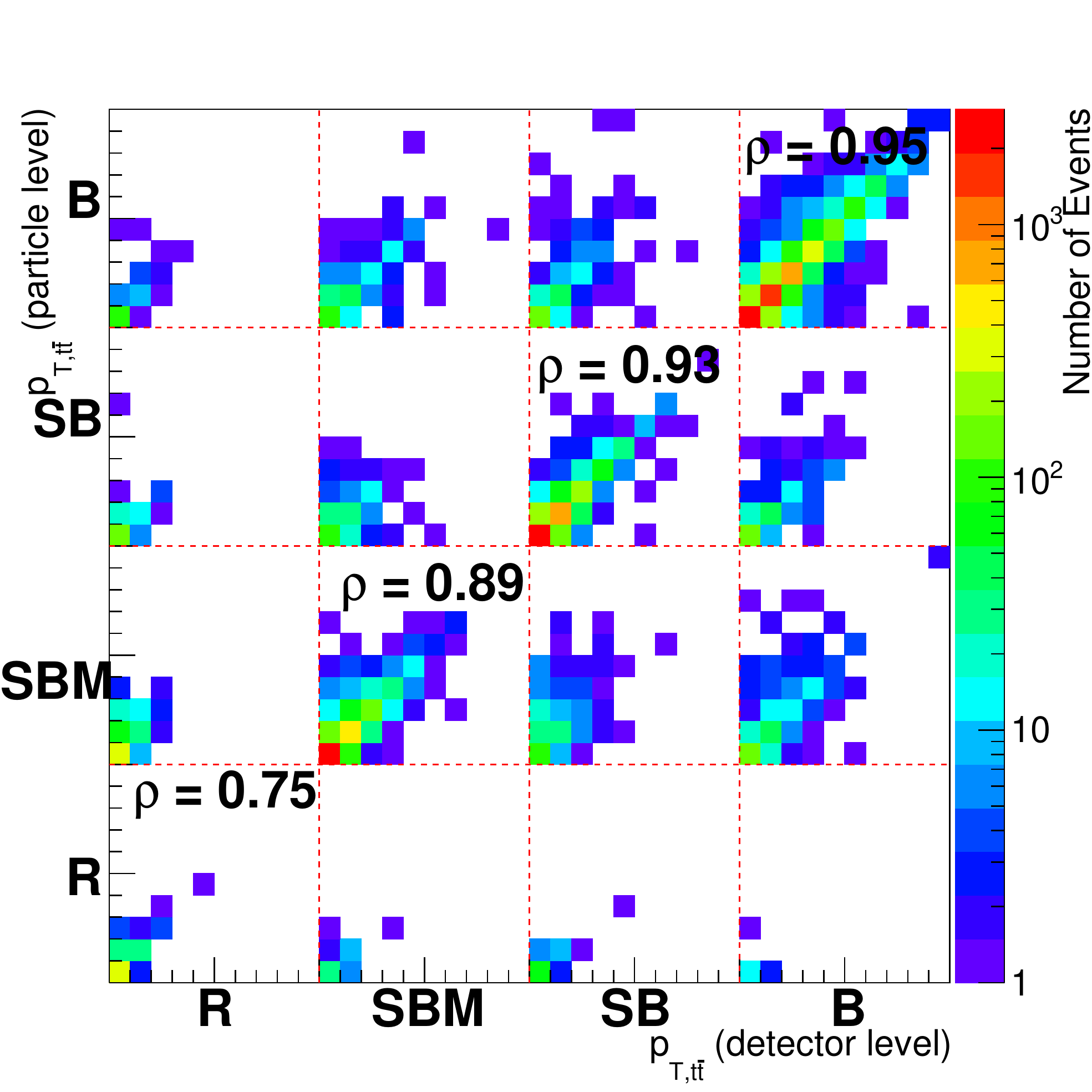}\\
\end{tabular}
\caption{The migration of events between the detector and the particle levels between the topologies (left) and an example of the migration of events for the transverse momentum of the $t\bar{t}$ system $p_{\mathrm{T,t\bar{t}}}$ between the resolved (R), semi-boosted mixed (SBM), semi-boosted (SB) and boosted (B) topologies (right) for the sample with $M_{\mathrm{y_{0}}} = 700$~GeV. The bin range for each of the sub-spectrum is 0--1000 GeV. }
\label{topo_mig}
\end{center}
\end{figure}

The kinematic reconstruction at the detector and particle levels is accomplished in parallel under the same selection requirements but the resulting event topologies are not necessarily the same at the two reconstruction levels. This is described by a migration matrix between the two levels in terms of the topologies as illustrated in Fig.~\ref{topo_mig} (left). Similar migrations further apply to values and bins of any studied observable. The example for the reconstructed $\ttbar{}$ pair transverse momentum ($p_{\mathrm{T,t\bar{t}}}$) is shown in Fig.~\ref{topo_mig} (right), this plot also shows migration of events between different bins. A matching condition, requiring the same topology at the detector and particle levels, is applied for the purpose of unfolding and only those events are taken into account to study migration between bins in selected spectra in the subsequent unfolding procedure.

\section{Results}

The results are summarized in this section consisting of the resolution of different samples, the performance of the unfolding process for selected variables in dedicated stacked samples; and the significance studies of a Beyond the Standard Model (BSM) signal over the Standard Model background before and after unfolding.

\subsection{Resolution of the $\ttbar{}$ resonance mass peak}

\begin{figure}  
\begin{center}
\begin{tabular}{cc}
\includegraphics[width=0.49\textwidth]{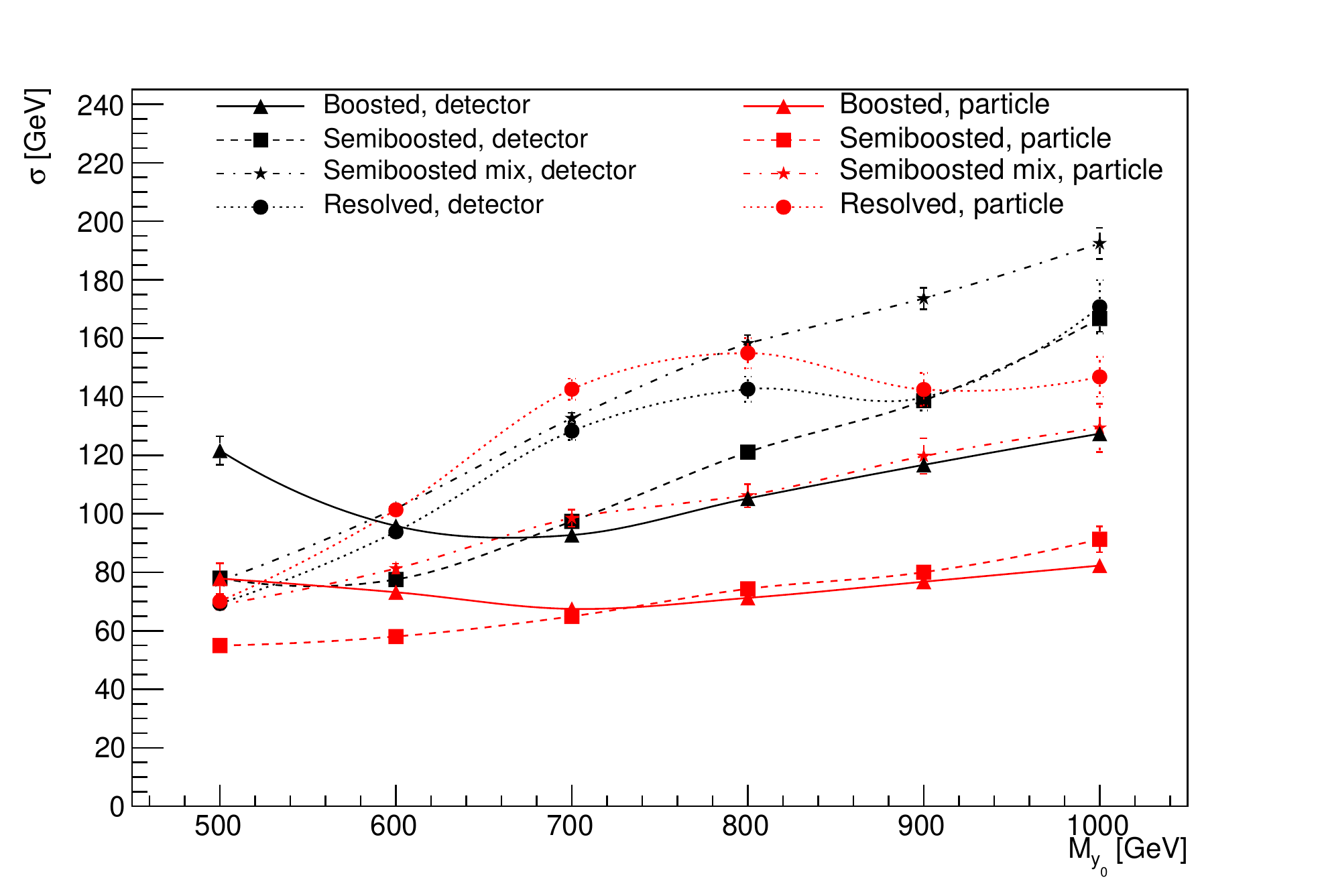} & 
\includegraphics[width=0.49\textwidth]{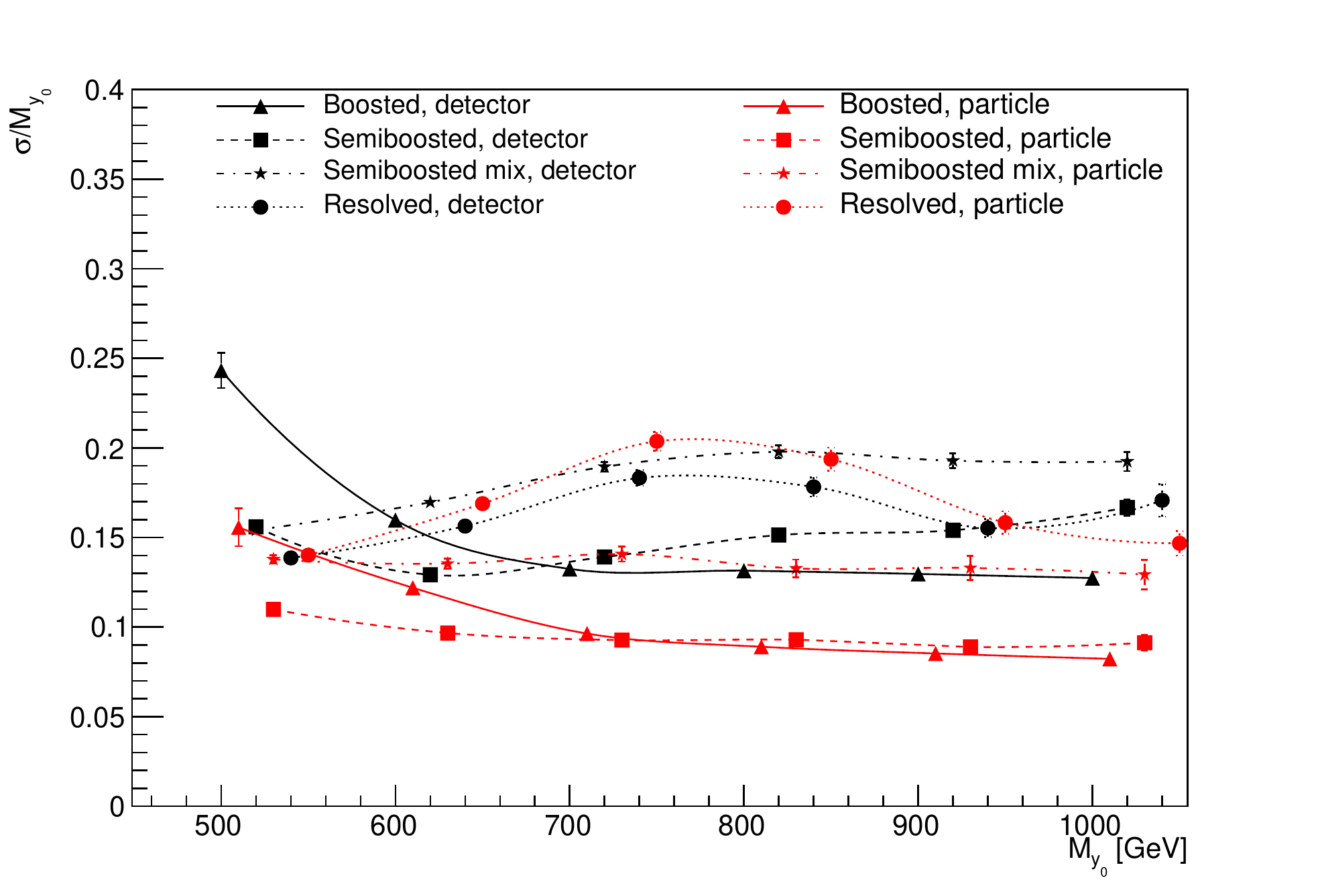}\\
\end{tabular}
\caption{The comparison of the $t\bar{t}$ system mass resolution for the samples with different masses of the hypothetical particle $y_{0}$ in the particular topologies the absolute (left) and relative (right) resolution with respect to $M_{\mathrm{y_{0}}}$. The horizontal shift in the position of markers around each mass point in the right plot is on the purpose to avoid the loss of information due to their overlap. }
\label{resolution}
\end{center}
\end{figure}

As expected for the $y_{\mathrm{0}}\rightarrow t\bar{t}$ BSM process, the reconstructed $t\bar{t}$ mass ($M_{\mathrm{t\bar{t}}}$) spectrum  peaks around the value of the mass of the studied hypothetical particle $y_{0}$ by construction. The width of the distribution is a~measure of the resolution of the $t\bar{t}$ resonance mass in given topology. A Gaussian curve was used to determine its width at both the detector and the particle levels and is shown in Fig.~\ref{resolution} as absolute (top) as well as relative (bottom), \textit{i.e.} when divided by the $y_0$ mass in the corresponding sample.

The relative resolution is comparable between all the topologies with the exception for the semi-boosted mixed topology at the detector level where the resolution is slightly worse, being approximately 15\% while the resolution of the other topologies is 9--13\%. 

\subsection{The unfolding procedure and significance tests}

The unfolding procedure corrects for finite detector resolution effects in the measured spectra. The Fully Bayesian Unfolding (FBU) \cite{choudalakis2012fully} was chosen for the purposes of this procedure as implemented with the PyMC3 package \cite{pymc3}. In short, FBU uses the Bayesian theorem to estimate the truth (here particle) level spectrum from the measured detector-level spectrum using the migration matrix. As such, the matrix must be evaluated using simulated events and is normalized so that it describes the migration of events in a selected spectrum from particle-level to detector level bins. FBU returns a binned posterior corresponding to the estimated probability distribution of the variable in each particle-level bin. Its maximum can be chosen as the truth-level estimate for the observable in given bin. Among the main advantages of this method are that the migration matrix does not need to be inverted (which is a numerically unstable task) nor the problems is regularized and therefore becoming modified as \textit{e.g.} in the singular value decomposition method \cite{H_cker_1996}. Further advantages are the absence of iterations (and a need to terminate them at some point) as in the iterative Bayesian unfolding \cite{dagostini2010improved}; and the full control over the result as the full probability density is revealed in each bin.

In general, the unfolding process can be described by the following formula
\begin{equation}
\hat{T}_{\mathrm{i}} = \frac{1}{f_{\mathrm{i,eff}}} M_{\mathrm{ij}}^{-1} f_{\mathrm{j,acc}} (D_{\mathrm{j}} - B_{\mathrm{j}})\,,
\end{equation}
\noindent where $\hat{T}_{i}$ is the estimate of the particle level spectrum in bin $i$, $f_{\mathrm{i,eff}}$ is the efficiency correction, $M_{\mathrm{ij}}^{-1}$ stands for the main unfolding procedure using the migration matrix\footnote{Fully Bayesian Unfolding does not use the inverted migration matrix for the estimation of the particle level spectra, the notation in the formula stands for a shorthand of the unfolding procedure in general.}~$M_{\mathrm{ij}}$, $f_{\mathrm{j,acc}}$ is the acceptance correction in detector level bin $j$, $D_{\mathrm{j}}$ is the measured detector level spectrum and $B_{\mathrm{j}}$ is the estimated background. The efficiency and acceptance correction factors are defined as

\begin{equation}
f_{\mathrm{i,eff}} = \frac{P_{\mathrm{t\bar{t},i}}^{\mathrm{match}}}{P_{\mathrm{t\bar{t},i}}}  \quad \mathrm{and} \quad       f_{\mathrm{j,acc}} = \frac{D_{\mathrm{t\bar{t},j}}^{\mathrm{match}}}{D_{\mathrm{t\bar{t},j}}}\,,
\end{equation} 

\noindent where, in the context of this study, $P_{\mathrm{t\bar{t},i}}$ is the particle-level spectrum in bin $i$ using the $t\bar{t}$ sample as the model process, while $P_{\mathrm{t\bar{t},i}}^{\mathrm{match}}$ is the particle level spectrum in bin $i$ for events matched to the detector level events, \emph{i.e.} only events reconstructed at both particle and detector levels in the same topology contribute to this spectrum. Similarly, $D_{\mathrm{t\bar{t},j}}$ is the detector-level spectrum in bin $j$ and $D_{\mathrm{t\bar{t},j}}^{\mathrm{match}}$ is the detector level spectrum in bin $j$ for events matched to the particle level. Both correction factors are in the range between 0 and 1 by definition and were prepared using a statistically independent $t\bar{t}$ sample w.r.t. the $\ttbar{}$ sample used to compose the spectra to be unfolded.

The test spectra entering the unfolding procedure have the addition of the $y_{0}$ signal sample with an amplified cross section by an ad hoc number of $10^{3}$ ($5\cdot 10^{3}$ in case of resolved topology) to study the impact of the unfolding on the strength of a well-present signal of similar significance in all topologies. 

\subsection{Unfolding selected spectra}
Unfolding of three selected spectra is described in this section on the example in the semi-boosted topology, namely the transverse momentum of the hadronically decaying top quark ($p_{\mathrm{T}}^{\mathrm{t,had}}$), the invariant mass of the reconstructed top anti-top quark pair ($M_{\mathrm{t\bar{t}}}$) and the production angle of the hadronically decaying top quark ($\cos\theta^{*}_{\mathrm{t,had}}$).

The simulated samples were divided into two statistically independent halves.
The sum of first such halves is taken as the pseudo-data serving as input into the unfolding procedure while their statistically independent counterparts are stacked to form the expected total prediction.
 
\begin{figure}  
\begin{center}
\begin{tabular}{cc}
\includegraphics[width=0.49\textwidth]{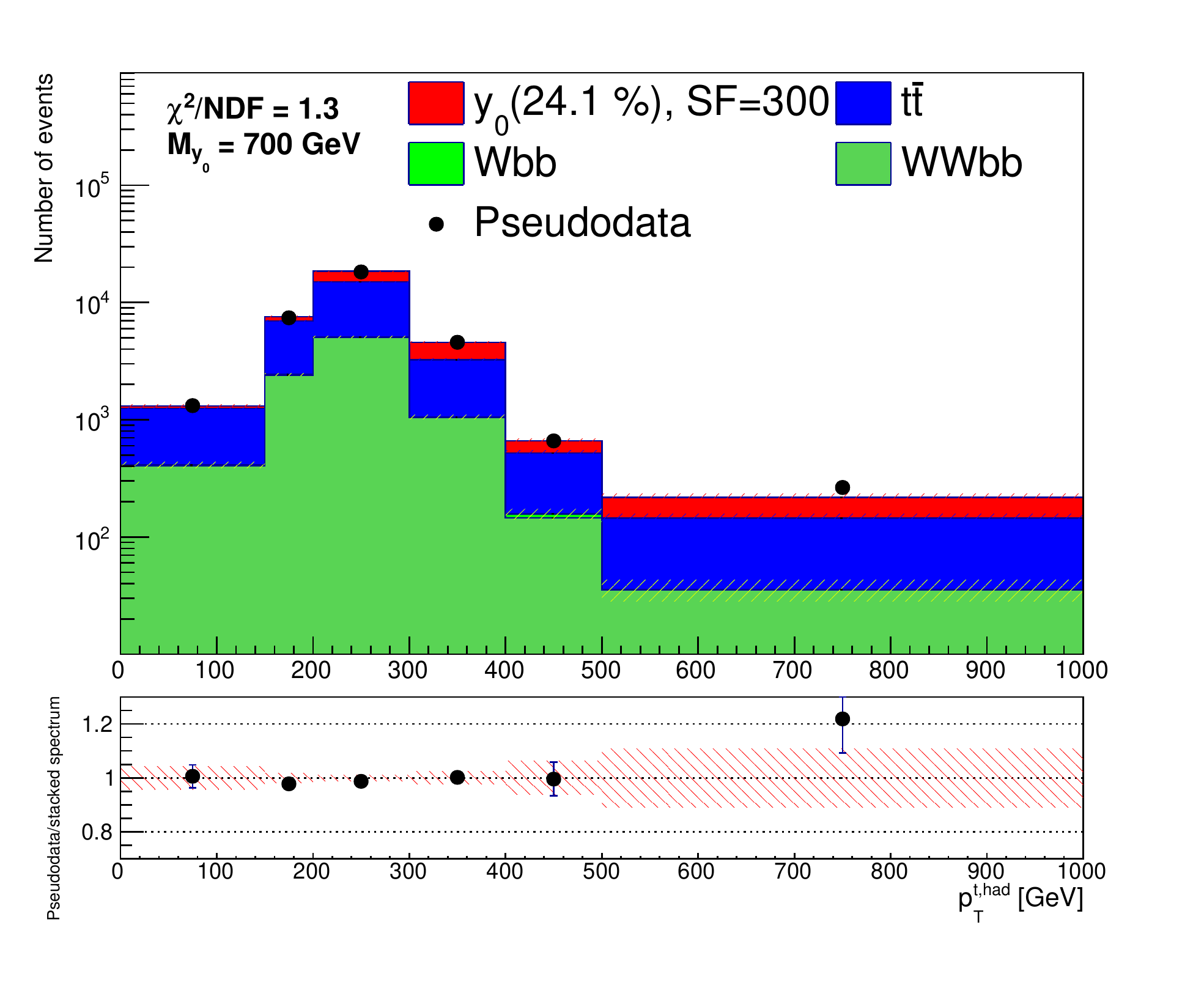} &
\includegraphics[width=0.49\textwidth]{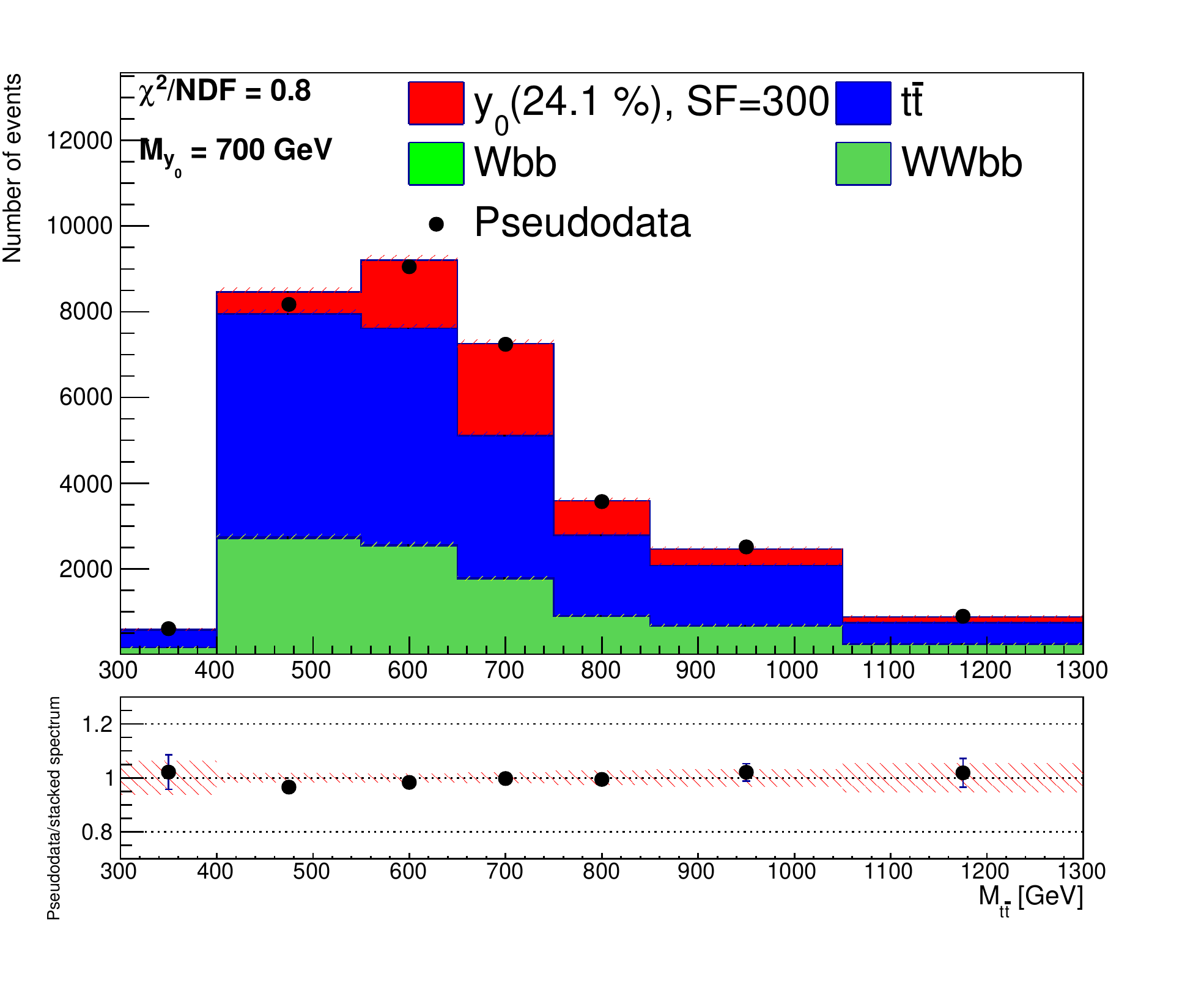} \\
\multicolumn{2}{c}{\includegraphics[width=0.49\textwidth]{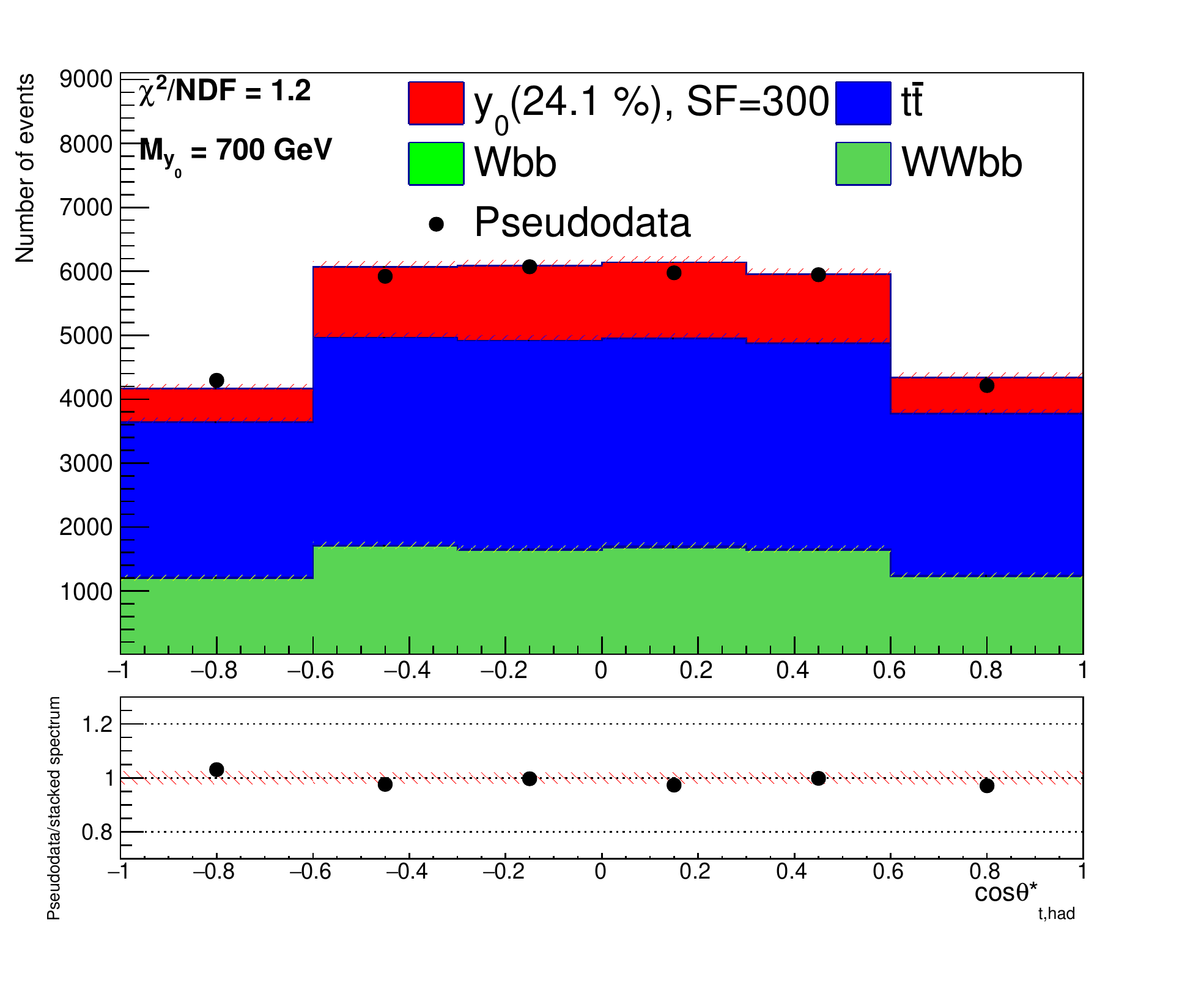}} \\
\end{tabular}
\caption{Comparison of the detector level spectra from two statistically independent parts (full markers and filled stack) for the $t\bar{t}$ sample with the addition of $Wbb$ and $WWbb$ backgrounds and an admixture of events from the sample with $M_{\mathrm{y_{0}}} = 700$ GeV for the transverse momentum of the hadronically decaying top quark ($p_{\mathrm{T}}^{\mathrm{t,had}}$, left), for the top anti-top quark pair invariant mass ($M_{\mathrm{t\bar{t}}}$, right) and for the crossing angle of the hadronically decaying top quark ($\cos\theta^{*}_{\mathrm{t,had}}$, bottom), all spectra are reconstructed in the semi-boosted topology. The hatched bands in the top panel represent the statistical uncertainty in each sample.}
\label{Unf_stacked}
\end{center}
\end{figure}

The comparison between the pseudo-data (black markers) and stacked histograms of the prediction (filled) for the transverse momentum of the hadronically decaying top quark in the semi-boosted topology is shown in Fig.~\ref{Unf_stacked} (left). The histogram and the stacked spectra agree well, their difference is within the statistical uncertainties. The binning was selected to respect the resolution, falling statistics, and so that the unfolding proceeds fast enough but still delivers useful information about spectra shape.   
The $t\bar{t}$ system invariant mass spectrum was chosen due to the possibility to see the hints of events from the sample with the hypothetical particle $y_{0}$. The spectrum entering the unfolding procedure and its statistically independent counterpart is shown in Fig.~\ref{Unf_stacked} (right).

The production angle of the hadronically decaying top quark was chosen to study the performance of this variable against the different topologies as an example of an angular variable. The general assumption is that boosted topology has jets preferably in smaller $|\eta|$ range due to the usually larger transverse momentum of the particles. This trend of the central $\eta$ preference drops in semi-boosted topologies and diminishes in the resolved topology. The spectrum used in unfolding and its statistically independent counterpart is also shown in Fig.~\ref{Unf_stacked} (bottom).

\begin{figure}  
\begin{center}
\begin{tabular}{cc}
\includegraphics[width=0.49\textwidth] {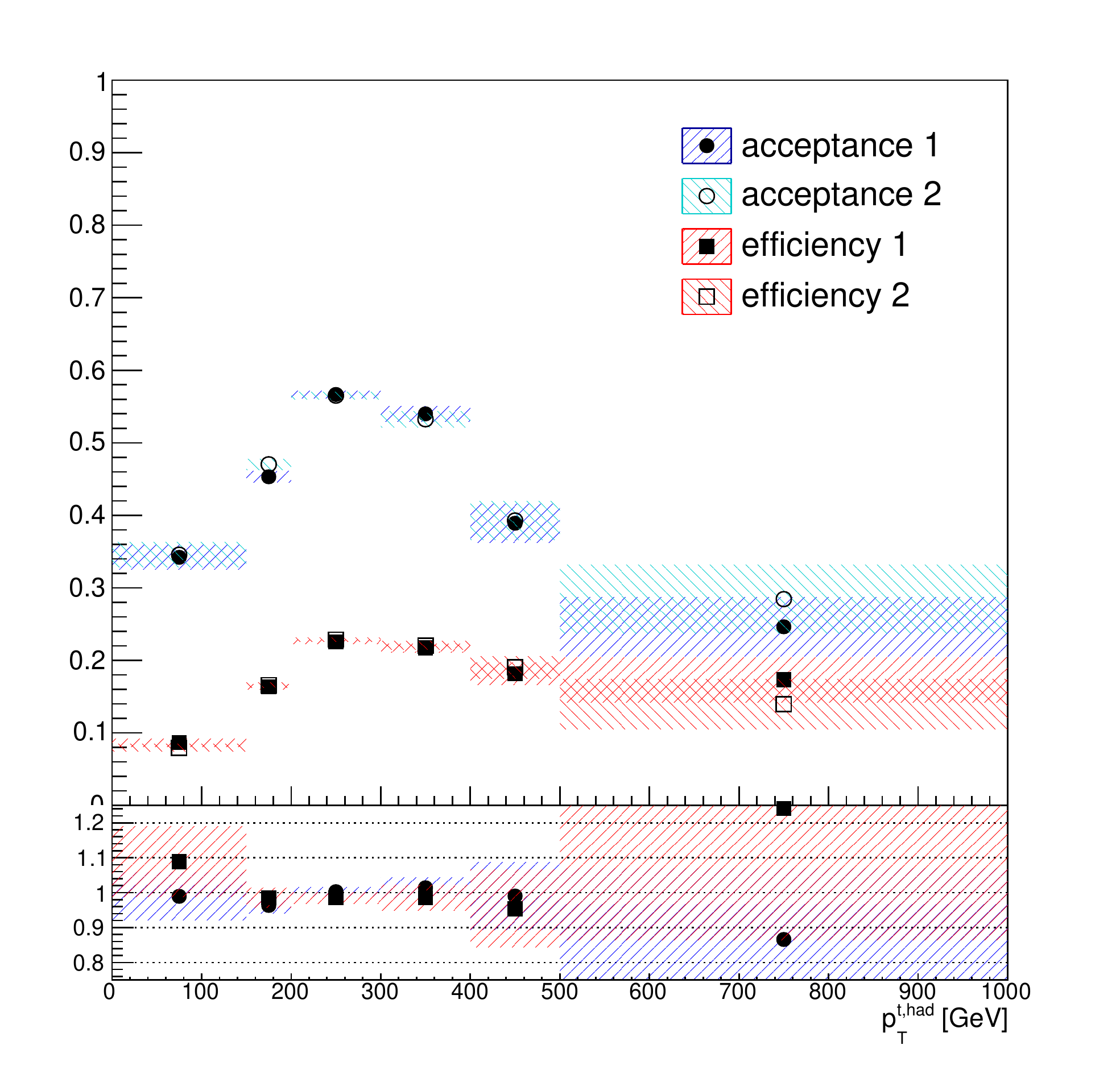} &
\includegraphics[width=0.49\textwidth] {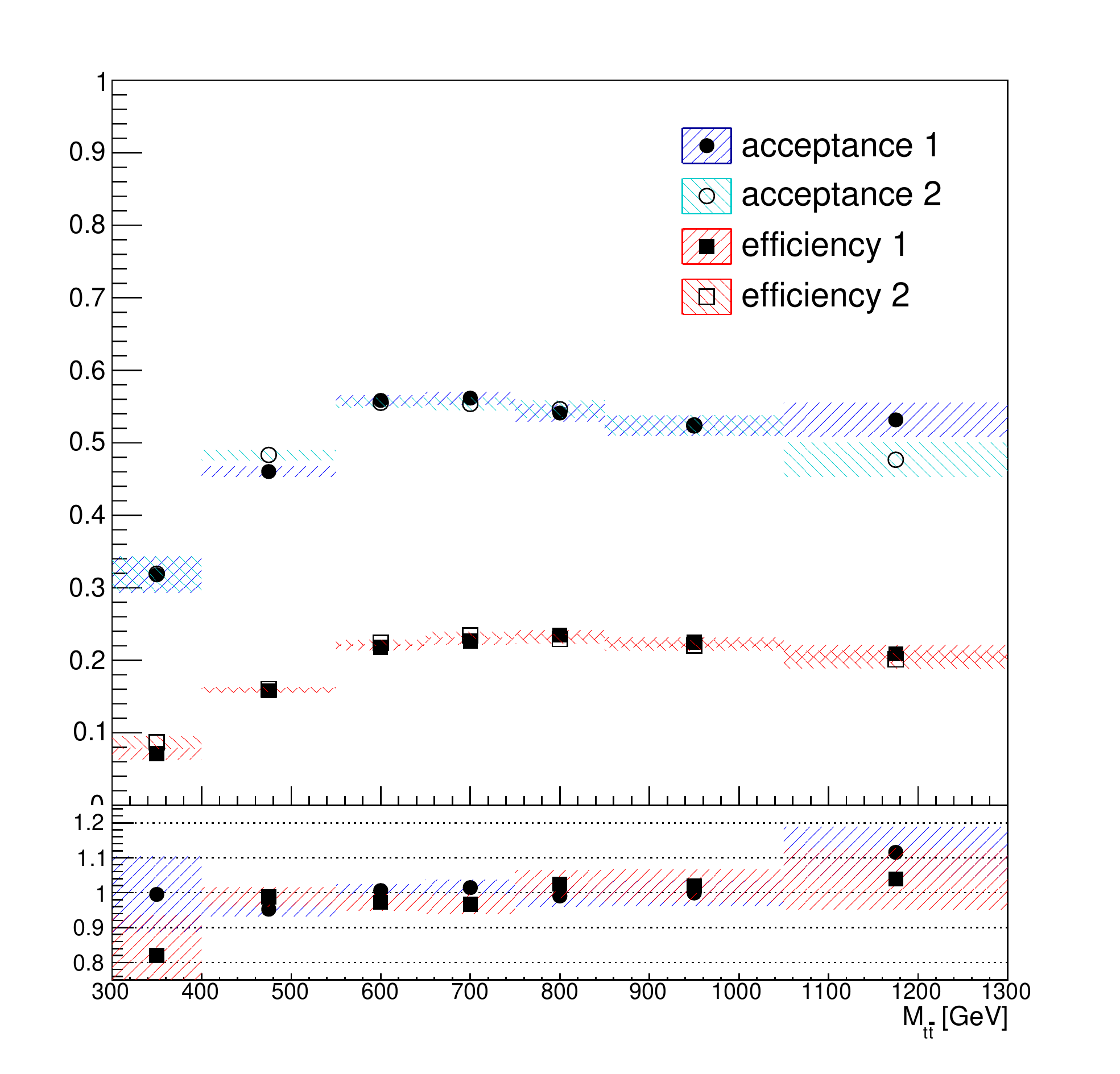} \\
\multicolumn{2}{c}{\includegraphics[width=0.49\textwidth] {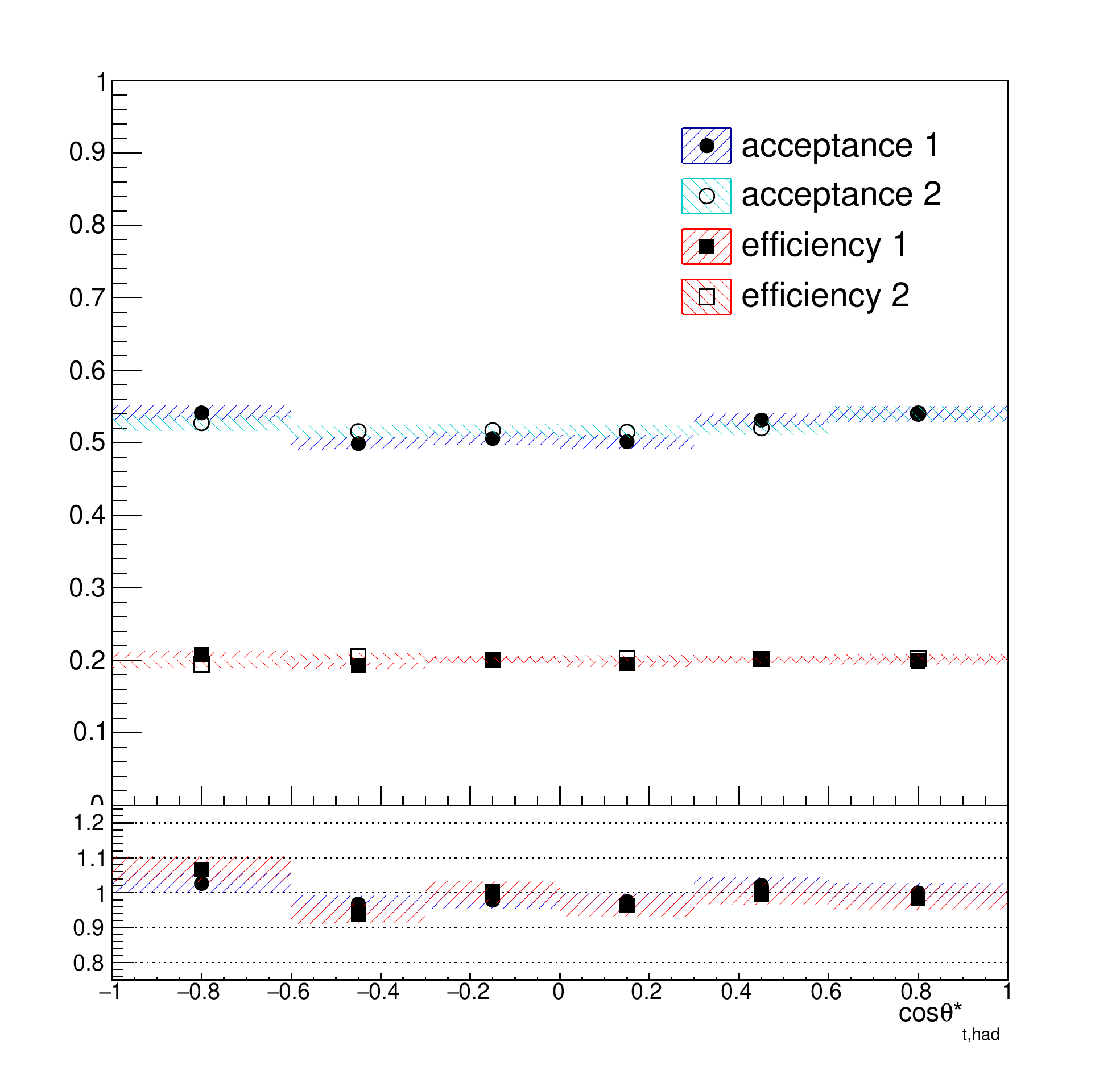}} \\
\end{tabular}
\caption{The acceptance and the efficiency for the two statistically independent $t\bar{t}$ samples for the reconstructed hadronically decaying top quark transverse momentum ($p_{\mathrm{T}}^{\mathrm{t,had}}$, left), the invariant mass of the reconstructed $t\bar{t}$ system ($M_{\mathrm{t\bar{t}}}$, right) and for the production angle of the hadronically decaying top quark ($\cos\theta^{*}_{\mathrm{t,had}}$, bottom). All spectra are reconstructed in the semi-boosted topology. Indices 1 and 2 denote the two statistically independent samples.}
\label{acc_eff_toph_pt}
\end{center}
\end{figure}

\begin{figure}  
\begin{center}
\begin{tabular}{cc}
\includegraphics[width=0.49\textwidth,valign=c]{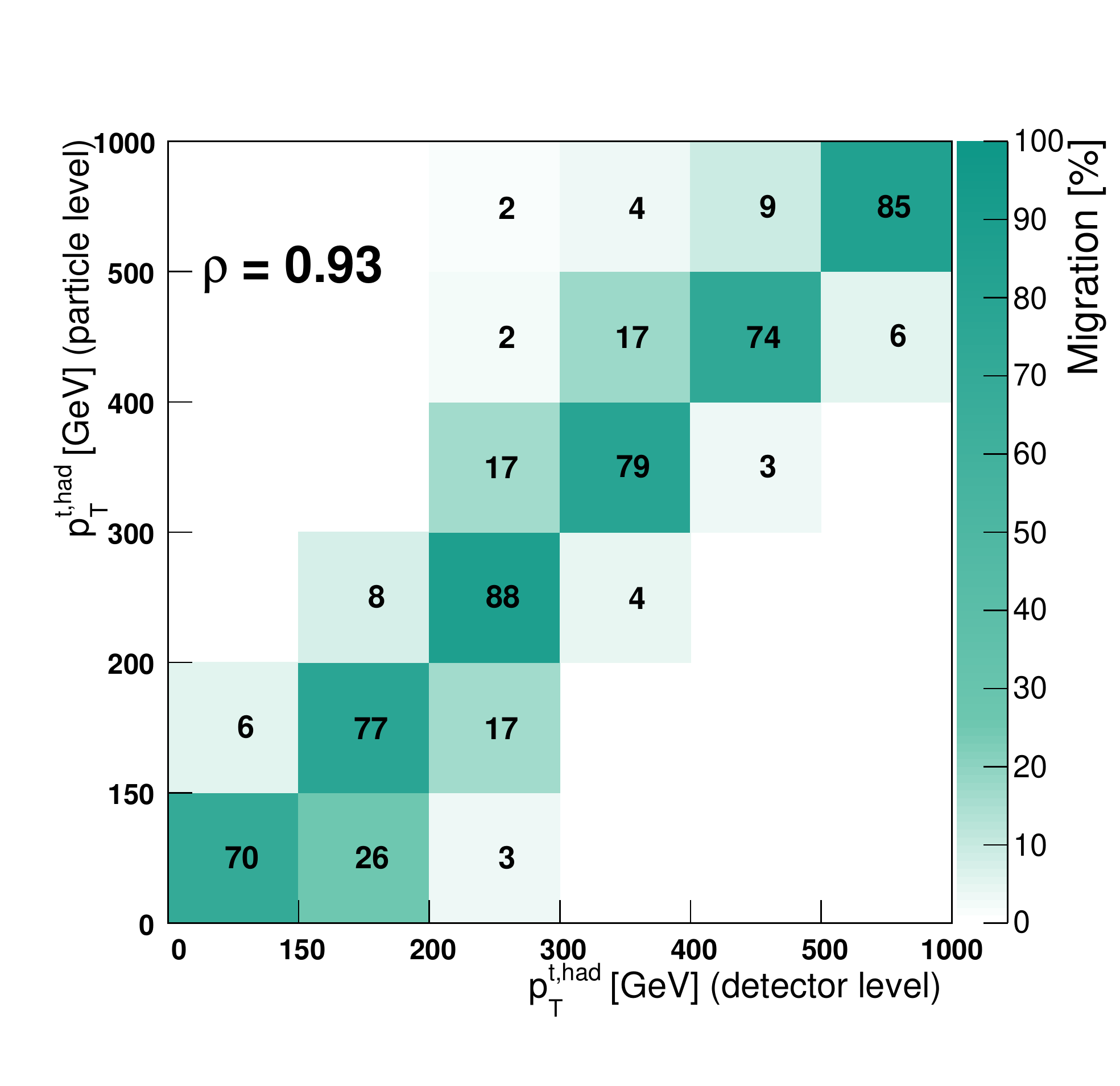} &
\includegraphics[width=0.49\textwidth,valign=c]{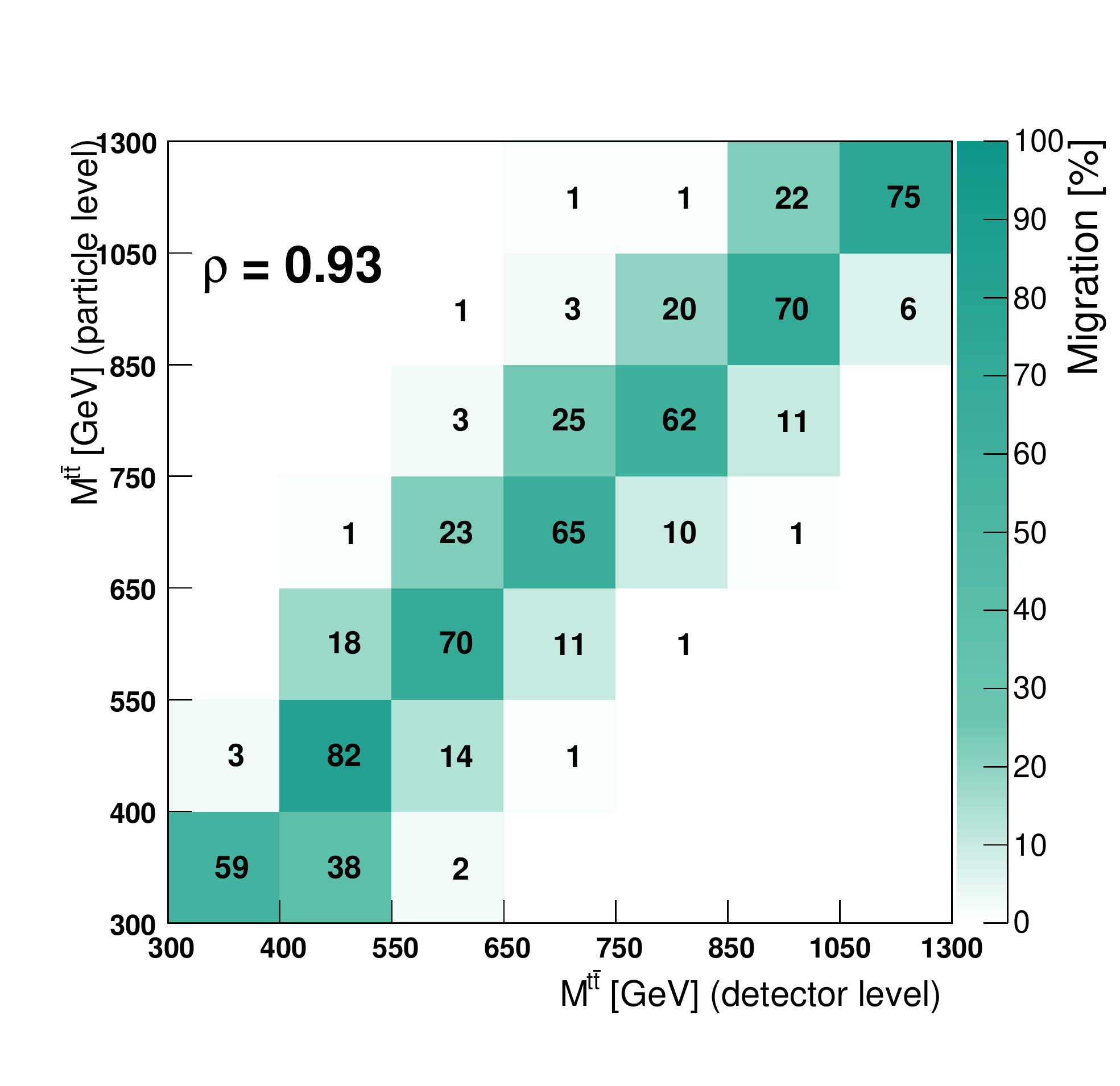} \\
\multicolumn{2}{c}{\includegraphics[width=0.49\textwidth, valign=c] {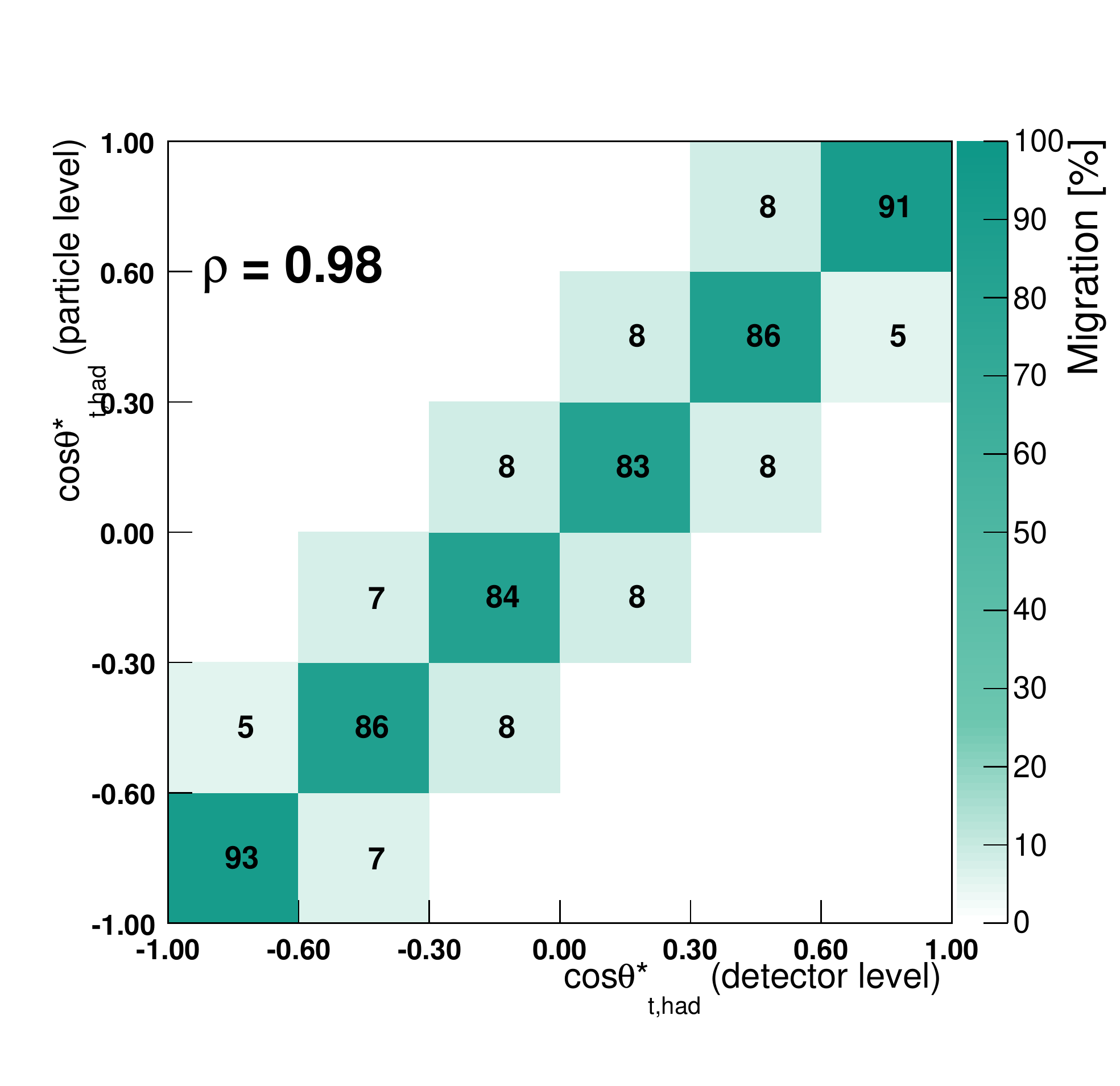}} \\
\end{tabular}
\caption{The migration matrices for the reconstructed transverse momentum of the hadronically decaying top quark ($p_{\mathrm{T}}^{\mathrm{t,had}}$, left) for the reconstructed invariant mass of $t\bar{t}$ system ($M_{\mathrm{t\bar{t}}}$, right) and for the crossing angle of the hadronically decaying top quark ($\cos\theta^{*}_{\mathrm{t,had}}$, bottom), all in the semi-boosted topology. All matrices were derived from the $t\bar{t}$ sample and used in the unfolding procedure. The correlation factor $\rho$ is calculated between the detector and particle levels.}
\label{matrices}
\end{center}
\end{figure}

The corrections used in the unfolding procedure were evaluated with the statistically independent sample from the one forming the pseudo-data using the $t\bar{t}$ sample. The corresponding acceptance and efficiency corrections are shown in Fig.~\ref{acc_eff_toph_pt} and the migration matrices are presented in Fig.~\ref{matrices}.

\begin{figure}  
\begin{center}
\begin{tabular}{ccc}
\includegraphics[width=0.49\textwidth]{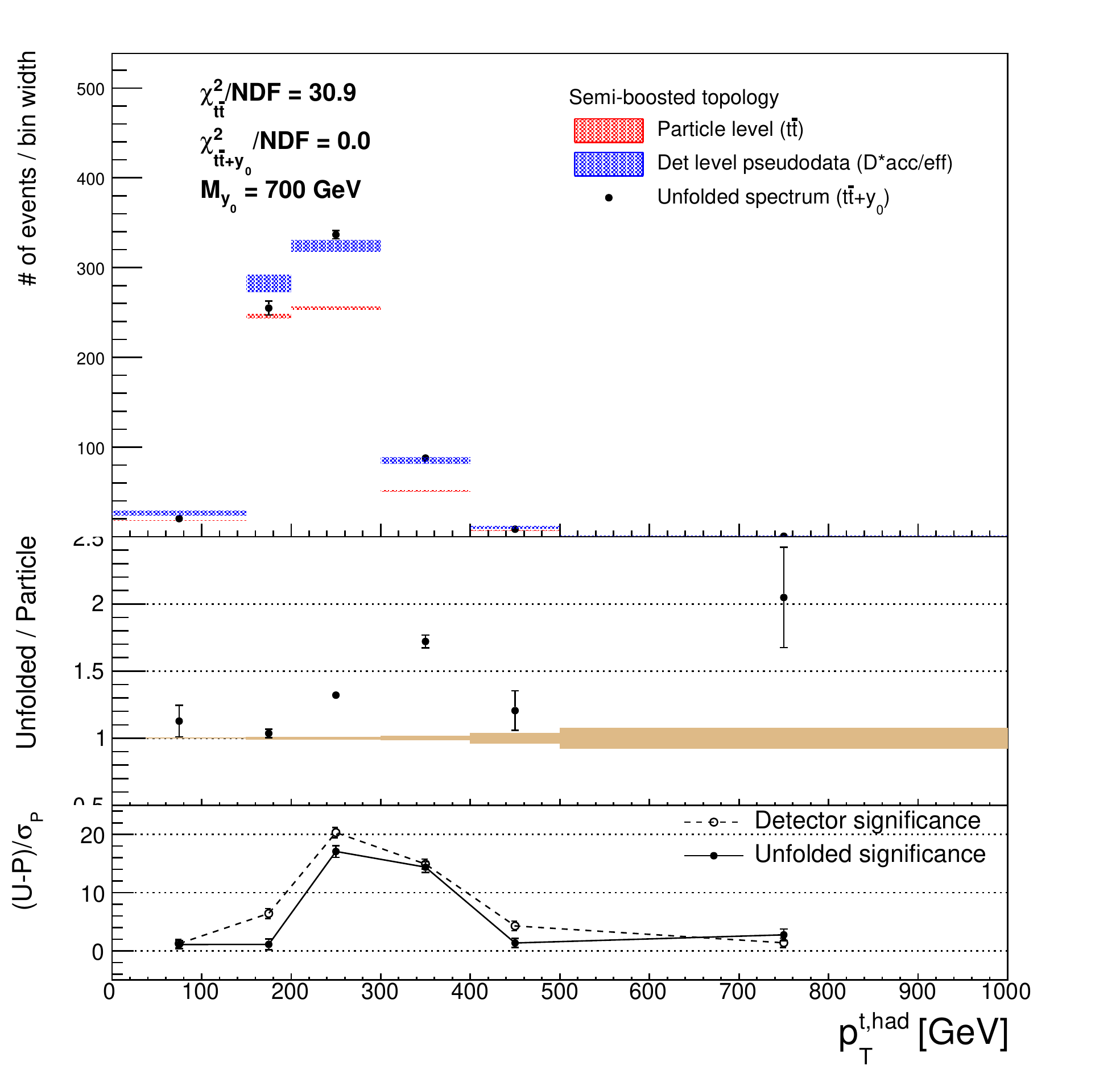} &
\includegraphics[width=0.49\textwidth]{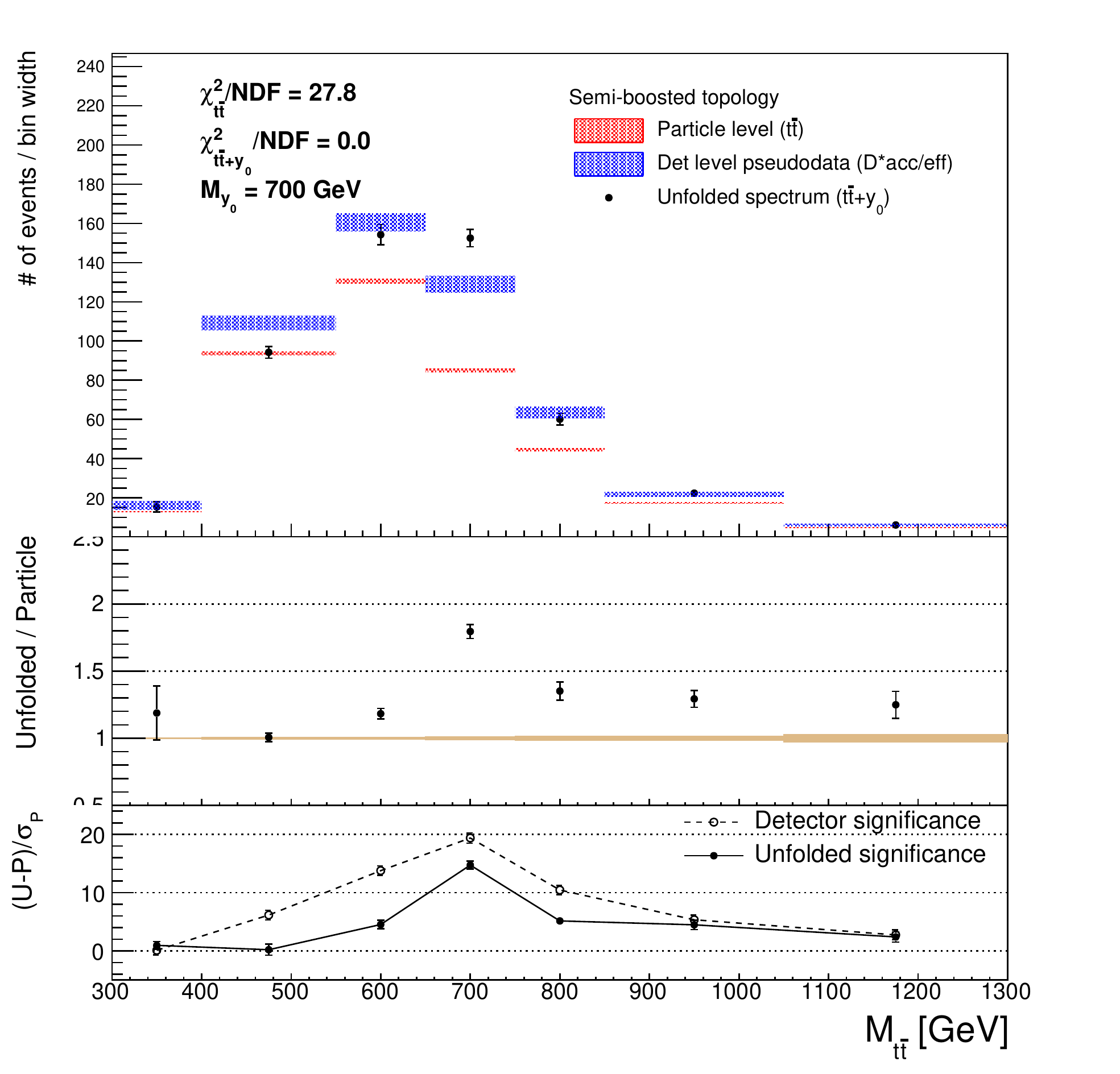} \\
\multicolumn{2}{c}{\includegraphics[width=0.49\textwidth] {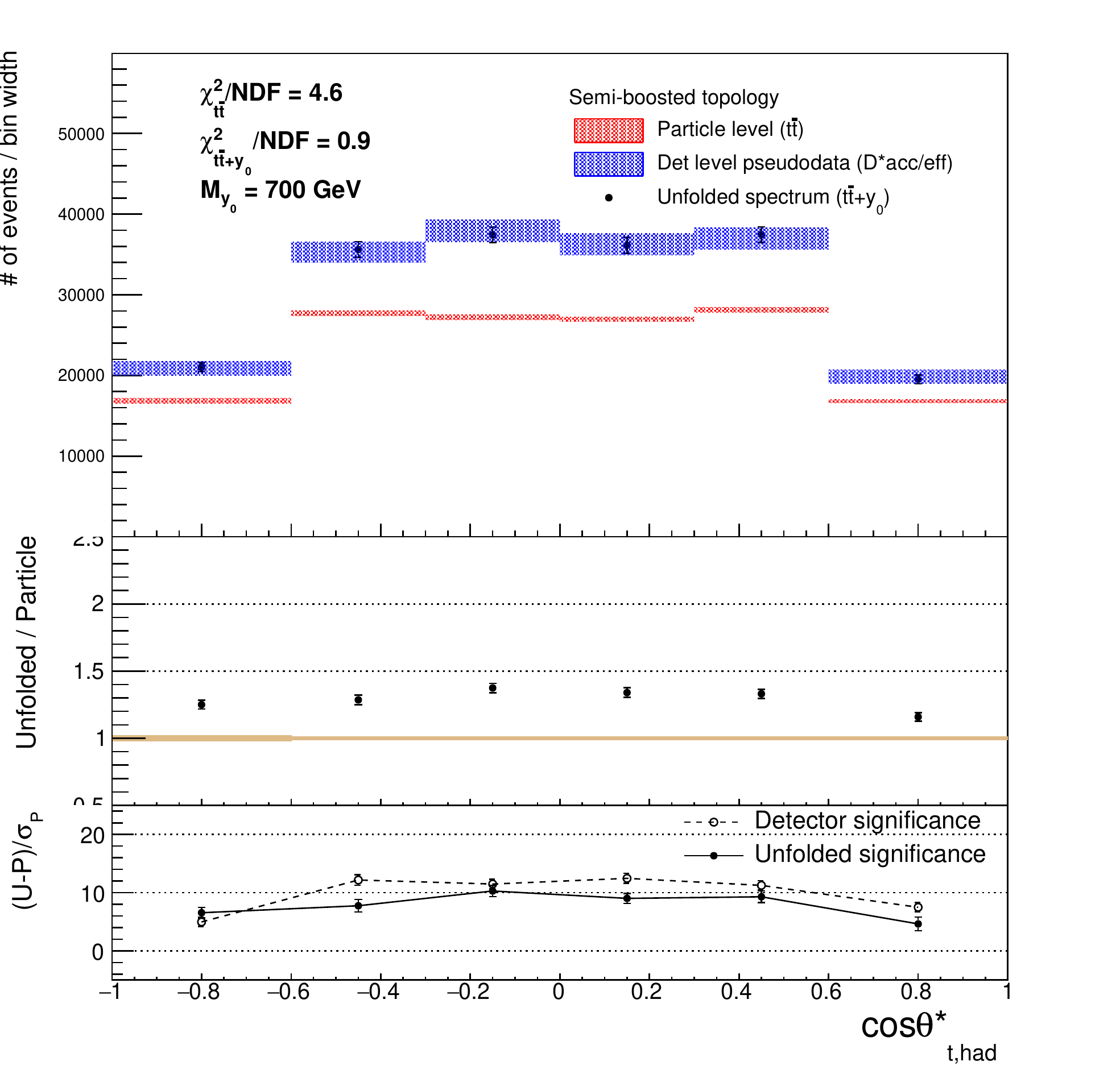}} \\
\end{tabular}
\caption{The comparison between the unfolded spectrum (black full markers), the detector level spectrum (blue) with the acceptance and efficiency correction applied to be comparable to the particle level one; and the $t\bar{t}$-only particle level spectrum (red) for the transverse momentum of the hadronically decaying top quark ($p_{\mathrm{T}}^{\mathrm{t,had}}$, top left), for the invariant mass of the reconstructed $t\bar{t}$ spectrum ($M_{\mathrm{t\bar{t}}}$, top right) and for the production angle of the hadronically decaying top quark ($\cos\theta^{*}_{\mathrm{t,had}}$, bottom), all in the semi-boosted topology. The $\chi^{2}_{\mathrm{t\bar{t}}}$ test is performed between the unfolded and $t\bar{t}$ particle level spectra while $\chi^{2}_{\mathrm{t\bar{t}+y_{0}}}$ between the unfolded and the $t\bar{t}$ particle level spectra with the $y_{0}$ signal included (closure test). The middle panels show the ratio of the unfolded spectrum over the particle level spectrum of the $t\bar{t}$ sample (full markers). Here the disagreement with the $t\bar{t}$-only particle level spectrum is caused by the addition of the $y_{0}$ signal before unfolding, the yellow band shows the statistical uncertainty in the particle level spectrum from the $t\bar{t}$ sample. The bottom panels show the detector (open markers, dashed line) and unfolded (full markers, solid line) $y_{0}$ signal significances in each bin. The detector-level $y_0$ signal significance is calculated using the original detector-level spectrum, \emph{i.e.} without applying the acceptance and efficiency correction.}
\label{unfolded_sb_pt}
\end{center}
\end{figure}

The unfolding results for the three spectra are shown in Fig.~\ref{unfolded_sb_pt}. A $\chi^2$ test was computed between the unfolded and the $t\bar{t}$ particle level spectra with the $y_{0}$ signal included, resulting in $\chi^{2}_{\mathrm{t\bar{t}+y_{0}}}$. As the input pseudo-data do contain the $y_0$ signal, this comparison is a closure test of the unfolding procedure ability to recover the particle level spectrum which consists of $\ttbar{}$ as well as the $y_0$ signal sample.
Another $\chi^{2}$ test was performed between the unfolded spectrum and the $t\bar{t}$-only particle level spectrum, resulting in $\chi^{2}_{\mathrm{t\bar{t}}}$. This tests evaluates the incompatibility of the unfolded pseudo-data with the $\ttbar$-only hypothesis.
The middle panels of Fig.~\ref{unfolded_sb_pt} show the ratio of the unfolded spectra over the particle level spectra from the $t\bar{t}$ sample (black full markers)where the disagreement with the $t\bar{t}$-only particle level spectrum is caused by the presence of the $y_{0}$ sample in the pseudo-data.

\subsection{Significance at the detector and unfolded levels}
The strength of the $y_{0}$ signal is quantified by the significance which considers the total statistical uncertainties in samples used in given bin. The significance $S$ in bin $i$ before unfolding is defined as
\begin{equation}
S_{\mathrm{i,det}} \equiv (P_{\mathrm{i}}^{t\bar{t}+y_{0}+B} - T_{\mathrm{i}}^{t\bar{t}}-B_{\mathrm{i},1} - \ldots - B_{\mathrm{i},k})/\sqrt{\sigma _{P_{\mathrm{i}}}^{2}+\sigma _{T_{\mathrm{i}}}^2 + \sigma _{B_{\mathrm{i},1}}^2 + \ldots}\,,
\end{equation}  

\noindent where $P_{\mathrm{i}}$ is the number of the the pseudo data events consisting from the signal and the background added to the expected $t\bar{t}$ sample in bin $i$; $T_{\mathrm{i}}$ is the detector level spectrum from the statistically independent $t\bar{t}$ sample, $B_{\mathrm{i},k}$ is the background contribution to the studied spectra from the $k$-th background sample; and $\sigma _{P_{\mathrm{i}}}$,  $\sigma _{T_{\mathrm{i}}}$  and  $\sigma _{B_{\mathrm{i},k}}$ are the statistical uncertainties in the pseudo data, $t\bar{t}$ and the $k$-th background samples, respectively (all at the detector level). The composition of the sample is denoted in the upper index, \emph{e.g.} $t\bar{t} + y_{0} + B$ stands for the sum of the background, $t\bar{t}$ and the $y_{0}$ samples. Systematic uncertainties are not part of this study as their effect would be largely coherent across topologies, thus not changing the conclusions nor hierarchy of the observed patterns.

A similar significance is defined after the unfolding procedure, before which the background was subtracted, as
\begin{equation}
S_{\mathrm{i,unf}} \equiv (U_{\mathrm{i}}^{t\bar{t}+y_{0}} - T_{\mathrm{i}}^{t\bar{t}})/\sqrt{\sigma _{U_{\mathrm{i}}}^2+\sigma _{T_{\mathrm{i}}}^2}\,,
\end{equation}  

\noindent where $U_{\mathrm{i}}$ is the number of the unfolded pseudo-data events in bin $i$, $T_{\mathrm{i}}$ is the particle level spectrum from the statistically independent $t\bar{t}$ sample; and $\sigma _{U_{\mathrm{i}}}$ and $\sigma _{T_{\mathrm{i}}}$ are the statistical uncertainties in the unfolded spectrum bin $i$ and in the statistically independent $t\bar{t}$ sample at the particle level, respectively. The binned detector and unfolded significances are shown under the ratio plots of the unfolded spectra in Fig.~\ref{unfolded_sb_pt}. The integral significance, strength of the signal over the whole spectrum, is defined similarly for both the detector and the unfolded level. The detector level integral significance is defined as
\begin{equation}
S_{\mathrm{I,det}} \equiv \sum_{i=0}^{m}( P_{\mathrm{i}}^{t\bar{t}+y_{0}+B} - T_{\mathrm{i}}^{t\bar{t}}-B_{\mathrm{i},1} - \ldots - B_{\mathrm{i},k})/\sqrt{\sum_{i=0}^{m}(\sigma _{P_{\mathrm{i}}}^{2}+\sigma _{T_{\mathrm{i}}}^2})\,,
\end{equation}
\noindent where $m$ is number of bins in given spectrum. The detector level integral significance is the same for all variables. 

\begin{figure}  
\begin{center}
\begin{tabular}{cc}
\includegraphics[width=0.49\textwidth]{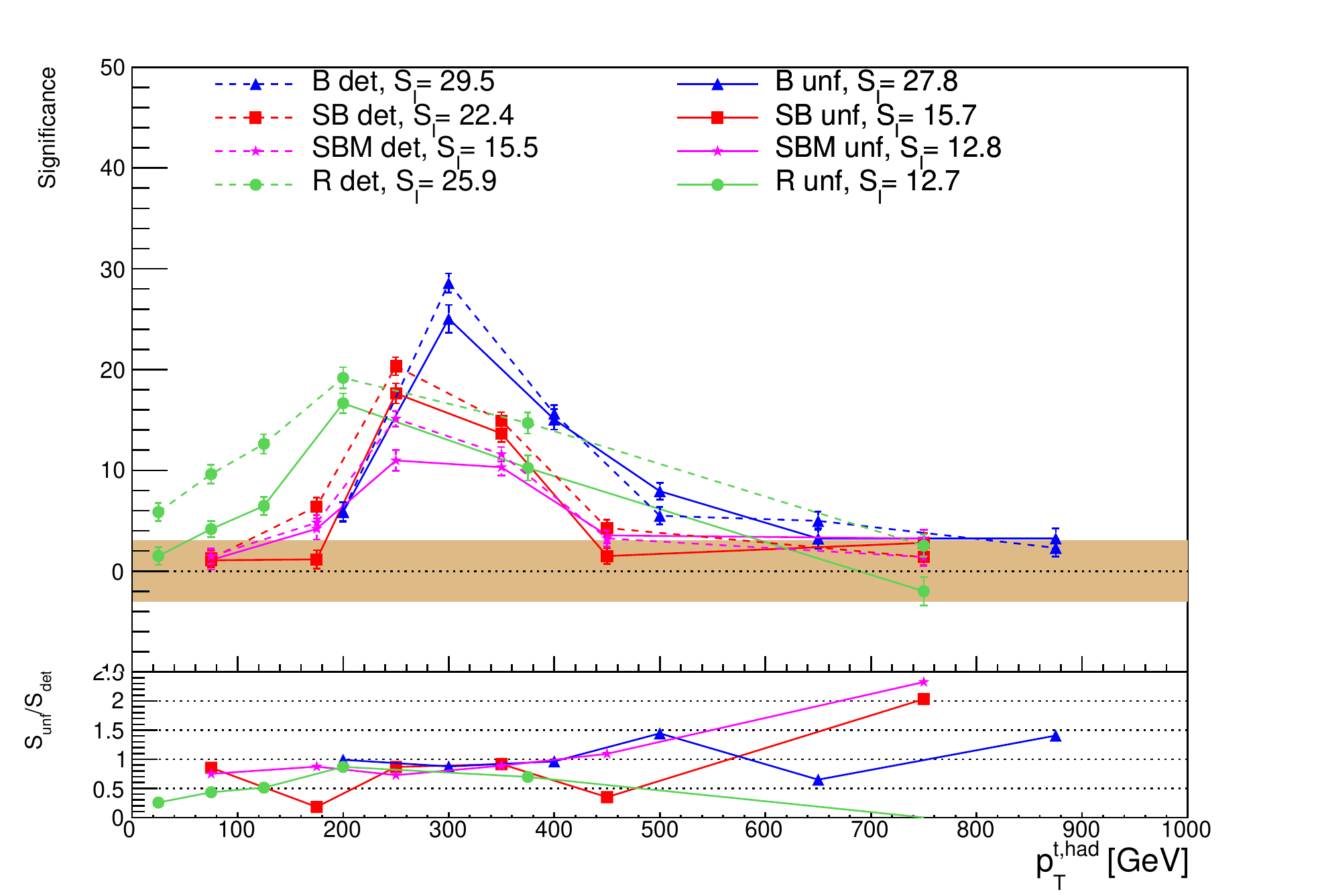}&
\includegraphics[width=0.49\textwidth]{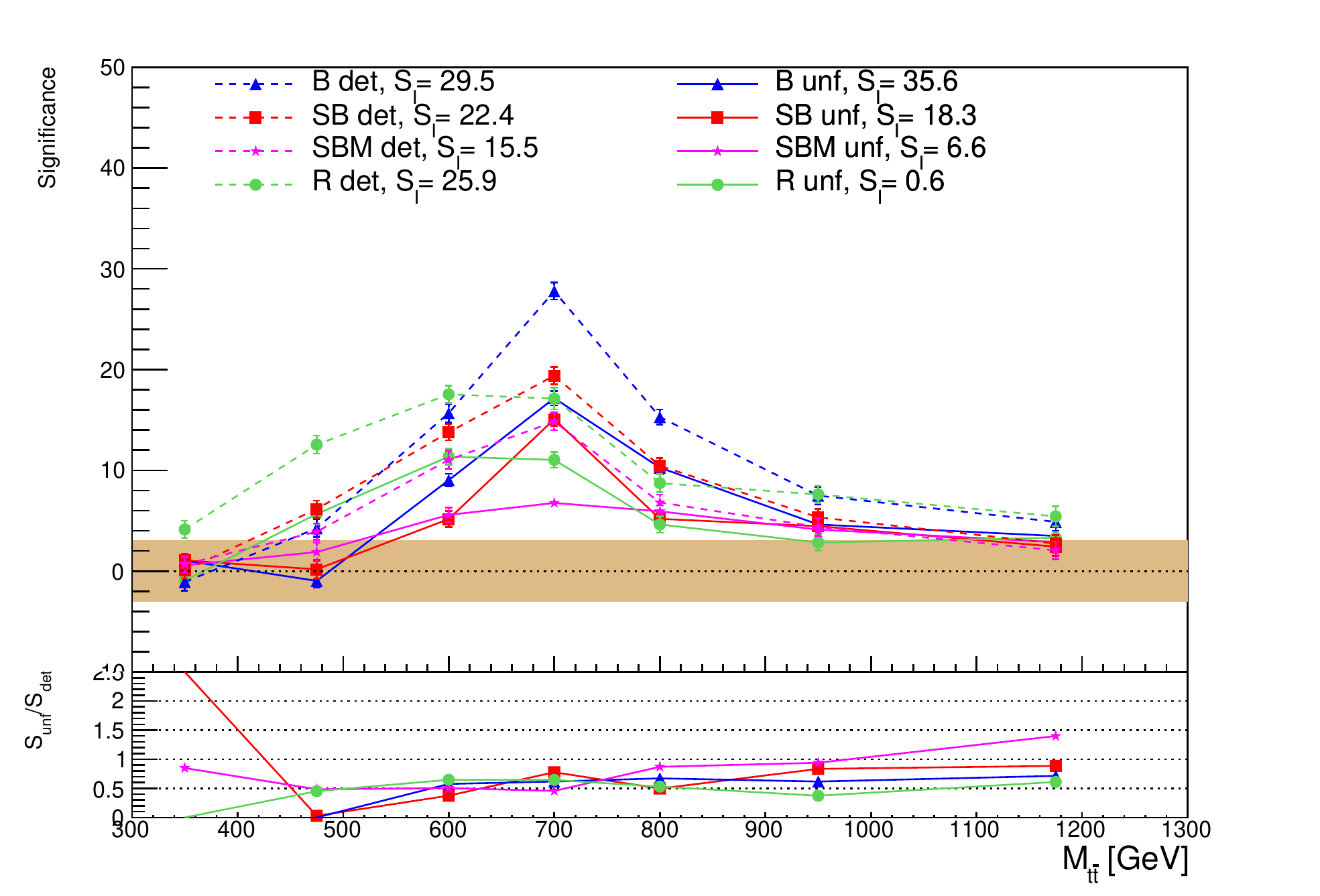}\\
\multicolumn{2}{c}{\includegraphics[width=0.49\textwidth]{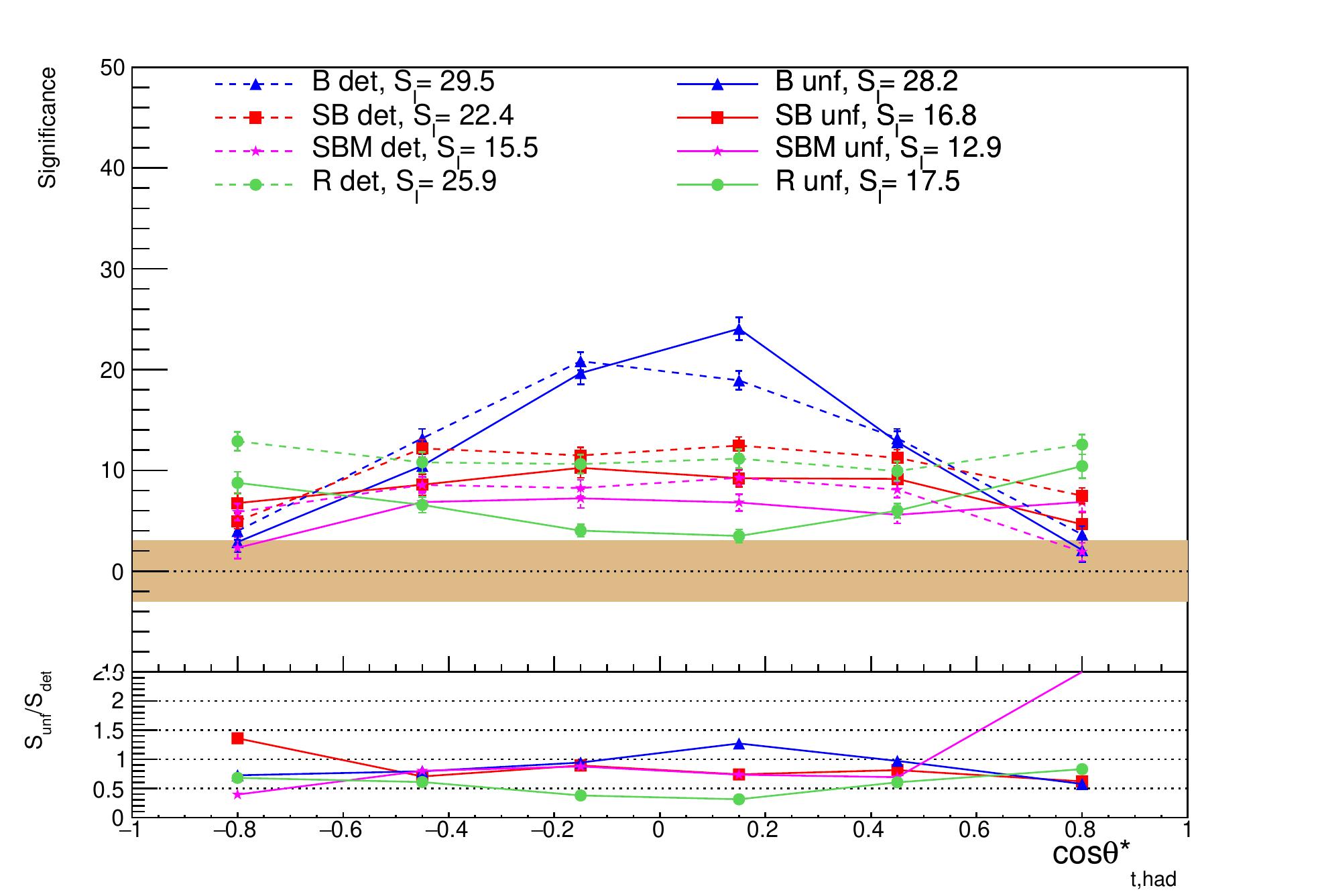}}\\
\end{tabular}
\caption{The detector (open markers, dashed line) and unfolded (full markers, solid line) significances for the transverse momentum of the hadronically decaying top quark ($p_{\mathrm{T}}^{\mathrm{t,had}}$, top left), invariant mass of the reconstructed $t\bar{t}$ pair ($M_{\mathrm{t\bar{t}}}$, top right) and the crossing angle of the hadronically decaying top quark ($\cos\theta^{*}_{\mathrm{t,had}}$, bottom) plotted for all the topologies in each bin. The orange band defines the area where the absolute value of the significance is below three, corresponding to the 3-$\sigma$ interval. The lower pads present the ratios of the unfolded over the detector level significances, without uncertainties which are highly correlated.}

\label{signs}
\end{center}
\end{figure}

\noindent The integral significance at the unfolded level is defined as
\begin{equation}
S_{\mathrm{I,unf}} \equiv \sum_{i=0}^{m}(U_{\mathrm{i}}^{t\bar{t}+y_{0}} - T_{\mathrm{i}}^{t\bar{t}})/\sqrt{\sum_{i=0}^{m}(\sigma _{U_{\mathrm{i}}}^2+\sigma _{T_{\mathrm{i}}}^2})\,.
\end{equation}  
\noindent The unfolded integral significance varies slightly over spectra as in the unfolding procedure the integral of the spectrum may not be preserved. The values of both the detector and the unfolded integral significance are presented in~the legend in Fig.~\ref{signs}.

The significances were calculated for the three selected spectra in all topologies and at both detector and particle levels. The comparison between significances before and after the unfolding procedure over the studied topologies are shown in~Fig.~\ref{signs} for the spectra of $p_{\mathrm{T}}^{\mathrm{t,had}}$, $M_{\mathrm{t\bar{t}}}$ and $\cos\theta^{*}_{\mathrm{t,had}}$.

The significance uncertainties were estimated by using 100 pseudo experiments for each spectrum with a smeared content in each detector-level bin. The smearing was performed by drawing a random number from the Gaussian distribution within the $\sigma$ parameter equal to the statistical uncertainty in the total detector-level spectrum in the given bin and with the mean parameter set to zero. Each such spectrum was unfolded using the same procedure and corrections and the binned significances were evaluated. The resulting standard deviation of significances in each bin is considered as the statistical uncertainty in the unfolded significance. The statistical uncertainty of the significances is already presented as error bars in~Fig.~\ref{signs}.

The signal significances in the $p_{\mathrm{T}}^{\mathrm{t,had}}$ spectrum peaks at different values of  $p_{\mathrm{T}}^{\mathrm{t,had}}$ depending on the topology as the event selection in each topology biases the spectrum and effectively selects different ranges in $M_{\mathrm{t\bar{t}}}$, too.
On the other hand, significance in the $M_{\mathrm{t\bar{t}}}$ spectrum peaks around the value of the generated $y_0$ mass of 700~GeV as expected, with a slight tail to lower values in the resolved topology which is the least suitable one to reconstruct a resonance of such a large mass.
In contrast, the $\cos\theta^{*}_{\mathrm{t,had}}$ is very flat also for the signal sample and there is no clear isolated excess of signal events in this spectrum, with the exception of the boosted topology which selects, by construction, high-$\pt$ large jets and thus also top quarks, consequently more localized in the central rapidity region, producing a pear around zero in $\cos\theta^{*}_{\mathrm{t,had}}$.

The three selected spectra are thus good candidate observables to illustrate different behavior and spread of significances over bins, also presenting a selection of observables of a dimension of energy as well as dimensionless (angular). The binned significances are in general lower after the unfolding, for which an explicit proof is delivered by this study. The cause of this is as follows.

While a sharper spectrum may be recovered by unfolding, the procedure in general correlates information among bins by maximizing a likelihood function in case of FBU, or minimizing (possibly modified and regularized) $\chi^2$ or iterating and sequentially improving the result for the case of other methods. An increase in the correlation across bins of the unfolded spectrum is a known and important fact and a correlation matrix should preferably be published along with unfolded spectra from real experiments, as done \emph{e.g.} in \cite{ATLAS:2014ipf,ATLAS:2017cez} where a correlation matrix between the observables was also evaluated. We observe that in case of the FBU method the posteriors variance usually increases, leading to larger absolute as well as relative uncertainties of the unfolded spectrum w.r.t. the particle level one. This effectively decreases the significance of the observed signal excess. An increase of the statistical uncertainties with the number of iterations in case of the Iterative Bayesian Unfolding~\cite{dagostini2010improved} was also reported by other analyses~\cite{ATLAS:2015dbj}. We note that in our case the BSM signal significances decrease by 20--40\%.
Other more explicit regularization methods like the SVD~\cite{H_cker_1996} provide a spectrum with a smaller statistical uncertainty by definition (effectively ditching small regularized response matrix eigenvalues which would lead to large variations), but are prone to unfolding biases towards the underlying simulation spectrum. Also the FBU extension with a regularization term leads to more narrow posteriors (smaller statistical uncertainty)~\cite{Baron:2020}. This places the standard FBU (without an explicit regularization term) among high-level unfolding methods with realistic statistical uncertainties.

\section{Conclusions}

The results of the semi-boosted and semi-boosted mixed reconstruction algorithm show potential to enhance the number of events in the $t\bar{t}$ analyses in the semi-leptonic decay channel. The estimates show the enrichment in events between 20\% and 50\% in the $t\bar{t}$ pair mass region ranging from 500 GeV to 1000 GeV. The resolution in the semi-boosted topology and the resolved or boosted topology is comparable, only the semi-boosted mixed has a worse resolution roughly by factor of 1.5.
The performance of the unfolding procedure including a simple background model shows results corresponding well to the particle level and the significance of the enhanced signal of the hypothetical $y_{0}$ particle is still visible after the unfolding. Values of the detector and the unfolded integral significance are comparable, yet there is 20--30\% decrease in significance between the detector and the unfolding levels caused by the unfolding in the reconstructed $t\bar{t}$ mass spectrum and 5--20\% in the reconstructed transverse momentum of the hadronically decaying top quark spectrum and the production angle of the top quarks, \textit{i.e.} both for energy-dependent as well as angular variables.

The concrete proof of diminishing a BSM signal significance by the unfolding procedure (using FBU) is, to our knowledge, shown explicitly for the first time in this study. Our findings thus support why the model-dependent searches for new physics at the LHC are mostly done at the detector level, using concrete detector-level (fully-simulated) BSM signals. We attribute the decrease of the significance to the increase of the statistical uncertainty in the unfolded spectrum, \emph{i.e.} to the posterior widening in the case of the FBU method. Due to this, although the $M_{\ttbar}$ spectrum becomes sharper after unfolding, revealing a narrower peak of the $y_0$ resonance, the binned significance does not increase due to larger statistical uncertainties of the unfolded spectrum caused by unfolding-induced correlation across bins. On the other hand, the significance stays substantial even after unfolding, opening doors to a comparison of any theory prediction at the particle-level to the unfolded data.

The described algorithm proves that semi-boosted and semi-boosted mixed topologies are sensitive to the possible presence of BSM signals. The selection criteria chosen close to those in real analyses make the studied algorithms applicable also in current LHC experiments.

\section{Acknowledgments}
The authors gratefully acknowledge the support from the Czech Science Foundation project GA\v{C}R 19-21484S and project IGA\_PrF\_2021\_004 of the Faculty of Science of the Palacky University Olomouc, Czech Republic. This work was performed as a part of fulfillment of doctoral studies of J. Pacalt at the Applied physics programme at the Faculty of Science of the Palacky University.

\bibliography{main}{}
\bibliographystyle{unsrt}
\newpage
\begin{appendices}
\section{Jet energy scale derivation and closure tests}
\label{appendix_JES}

The jet energy scale (JES) procedure corrects for a finite energy response of the detector to hadronic final states (jets) as function of their angle in the detector and energy. The goal is to correct detector-level jet energies to the particle level. The method to calculate the jet response thus uses information from Monte Carlo simulation and forms a ratio between value of the jet energy measured at the simulated detector $E_{\mathrm{det}}$ to the energy of a angularly matched the particle level jet ($E_{\mathrm{ptcl}}$).
In detail, first a jet at the detector level is chosen, then the particle level jet candidate with the smallest distance parameter defined as
\begin{equation}
 \Delta R = \sqrt{\Delta\eta^{2} + \Delta\phi^{2}}  < R_{\mathrm{cut}} \,.
\end{equation}
is chosen as the matching jet, with the $R_{\mathrm{cut}}$ parameter set to 0.2 for small ($R=0.4$) jets and 0.3 for large ($R=1$) jets.

\begin{figure}[!h]
\begin{center}
\begin{tabular}{cc}
\includegraphics[width=0.45\textwidth]{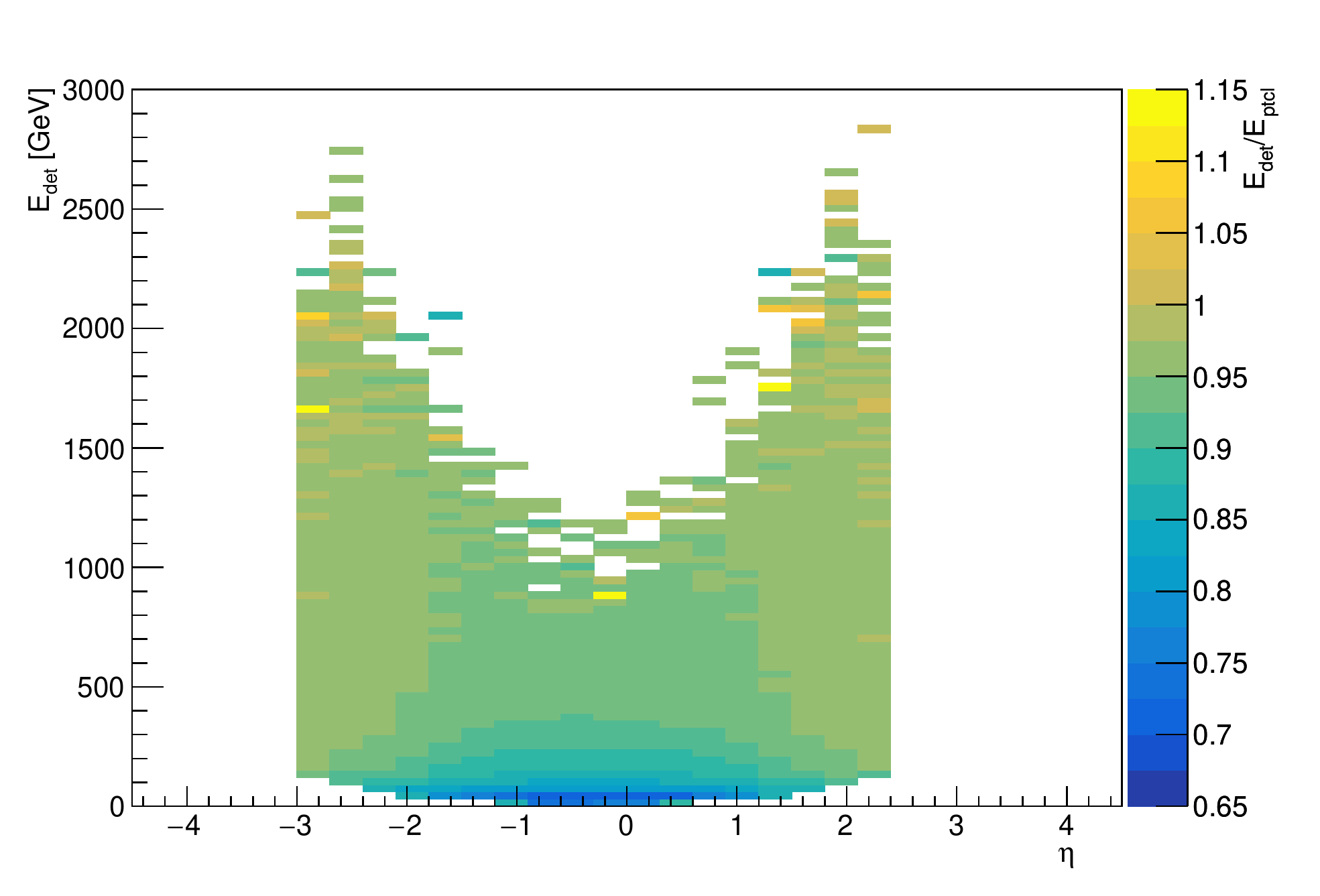} &
\includegraphics[width=0.45\textwidth]{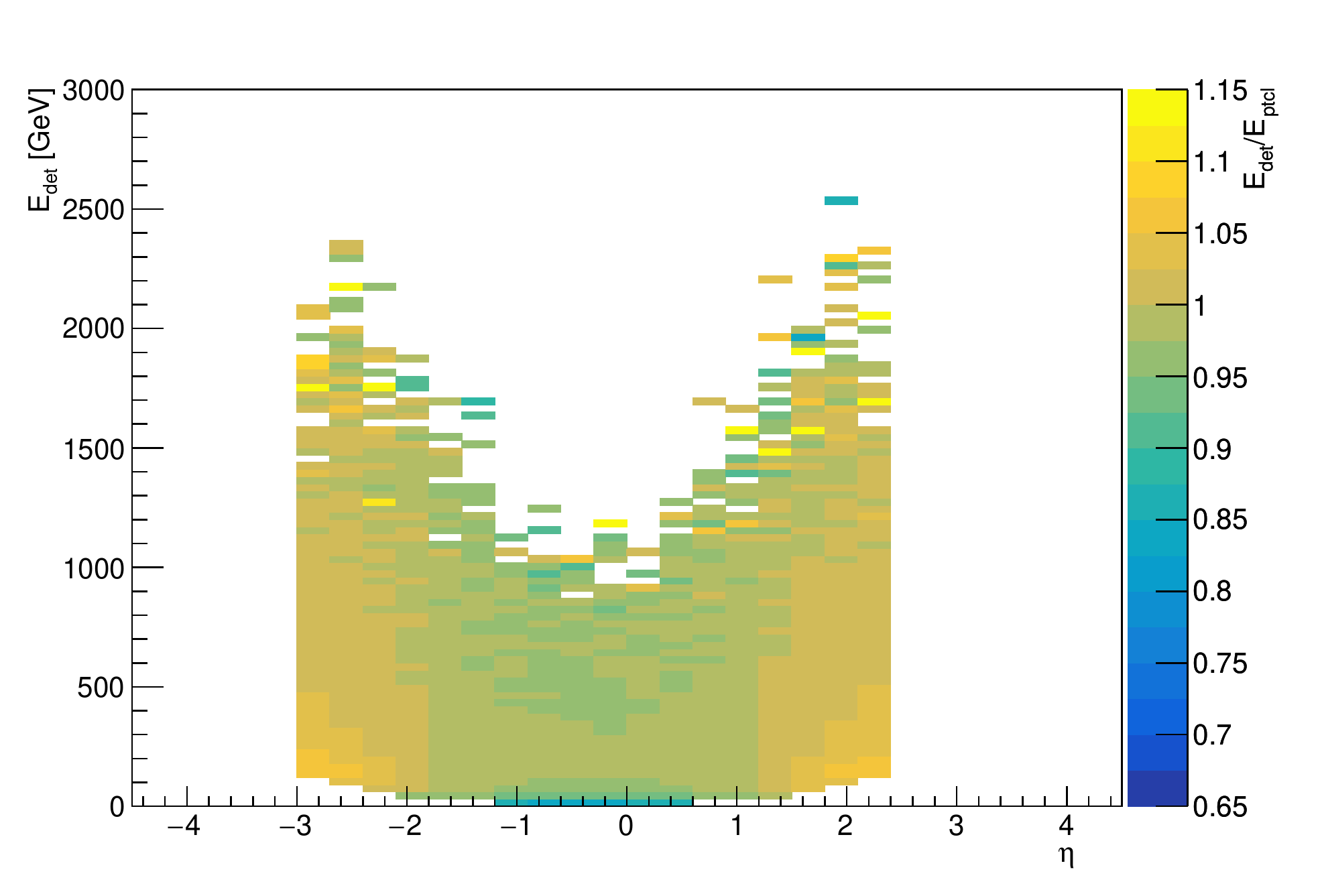} \\
\end{tabular}
\caption{The visualization of derived JES correction functions for large jets (left) and for small jets (right).}
\label{fig_JES_vis}
\end{center}
\end{figure}

The correction is binned in the detector-level jet pseudorapidity and transverse momentum 
\begin{equation}
\mathrm{JES}(\eta, p_{\mathrm{{T}}}) = \left\langle \frac{E_{\mathrm{ptcl}}^{\mathrm{jet}}}{E_{\mathrm{det}}^{\mathrm{jet}}} \right\rangle,
\end{equation}
and so the inverse value of the JES correction is the response of the detector-level jet.
Histograms of the jet response corresponding to different energy intervals are filled, each fitted
by a Gaussian function. The mean of the fit is plotted against the reconstructed energy and fitted by a polynomial logarithmic
function which is used to interpolate the JES correction to any energy. The visualization of the derived JES correction in dependence on the jet $\pt$ and $\eta$ is in Fig.~\ref{fig_JES_vis} for small jets (left) and large jets (right). 

A closure test was performed, in which the JES factors are applied to jets on a statistically independent sample to the one used to derive the JES corrections, but otherwise generated under the same settings.
The correction is re-derived on this already pre-corrected sample, thus the corrected detector level jet energy over the matched particle level jet energy is expected to be around unity by construction.

\begin{figure}[!h]
\begin{center}
\begin{tabular}{cc}
\includegraphics[width=0.45\textwidth]{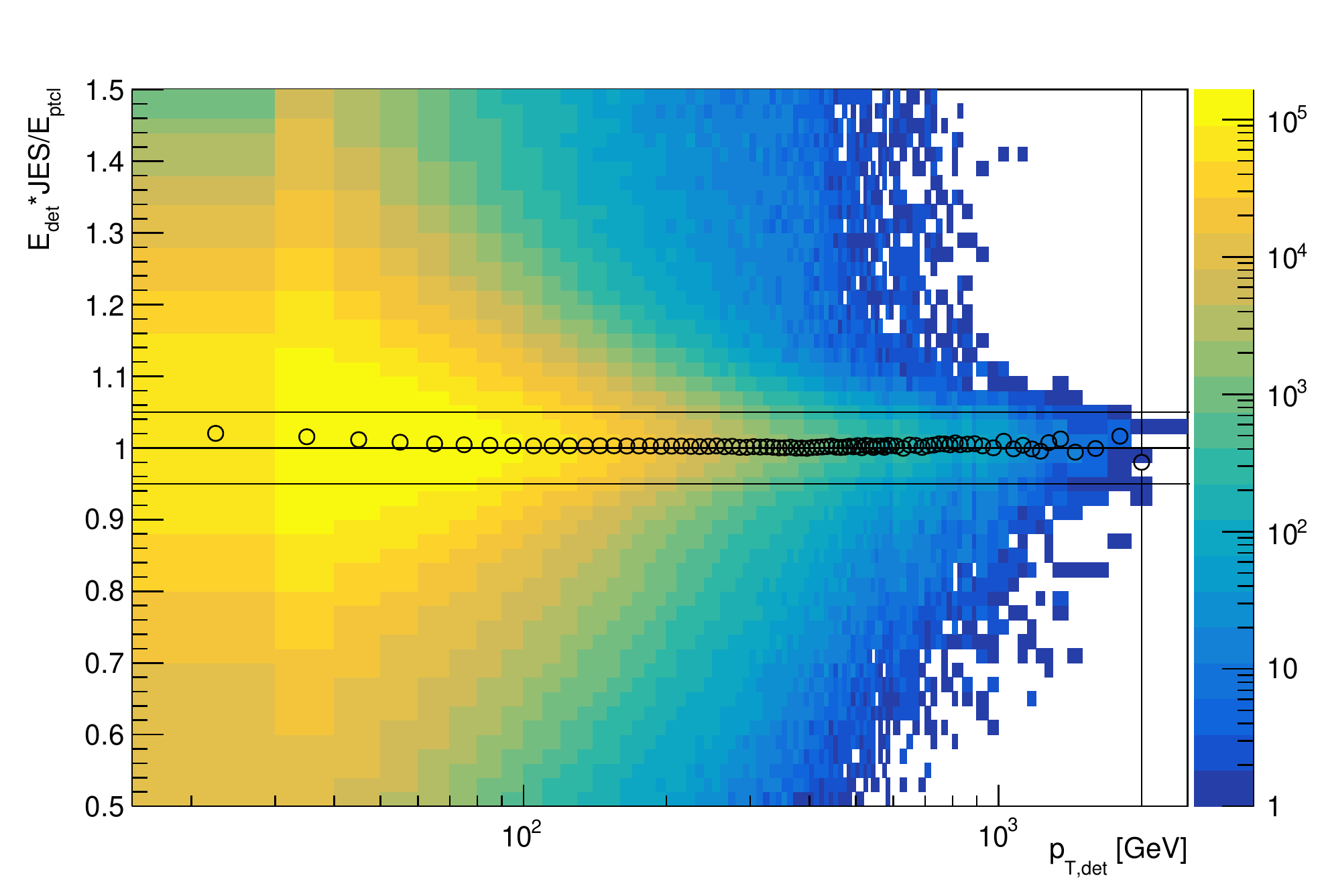} &
\includegraphics[width=0.45\textwidth]{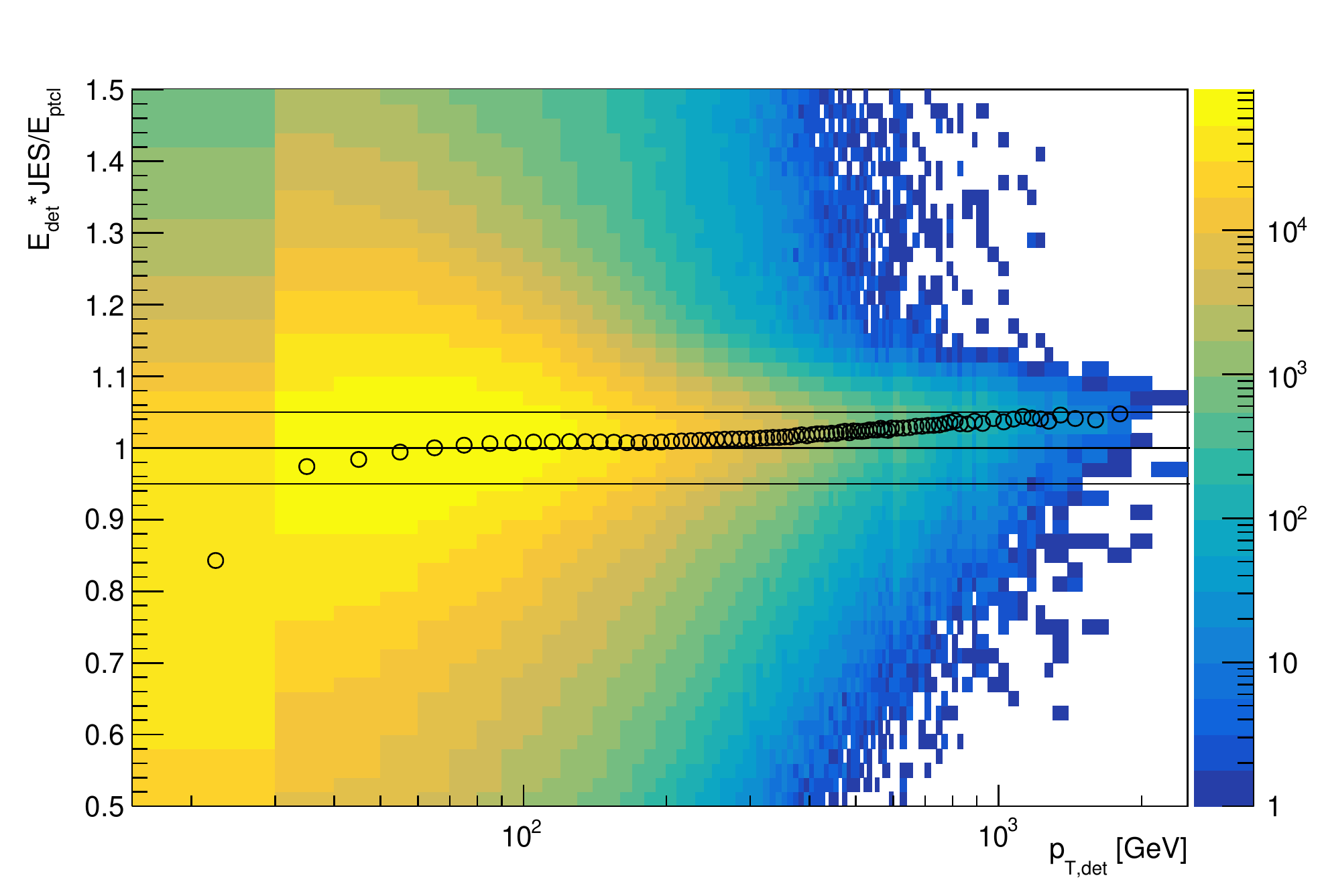} \\
\includegraphics[width=0.45\textwidth]{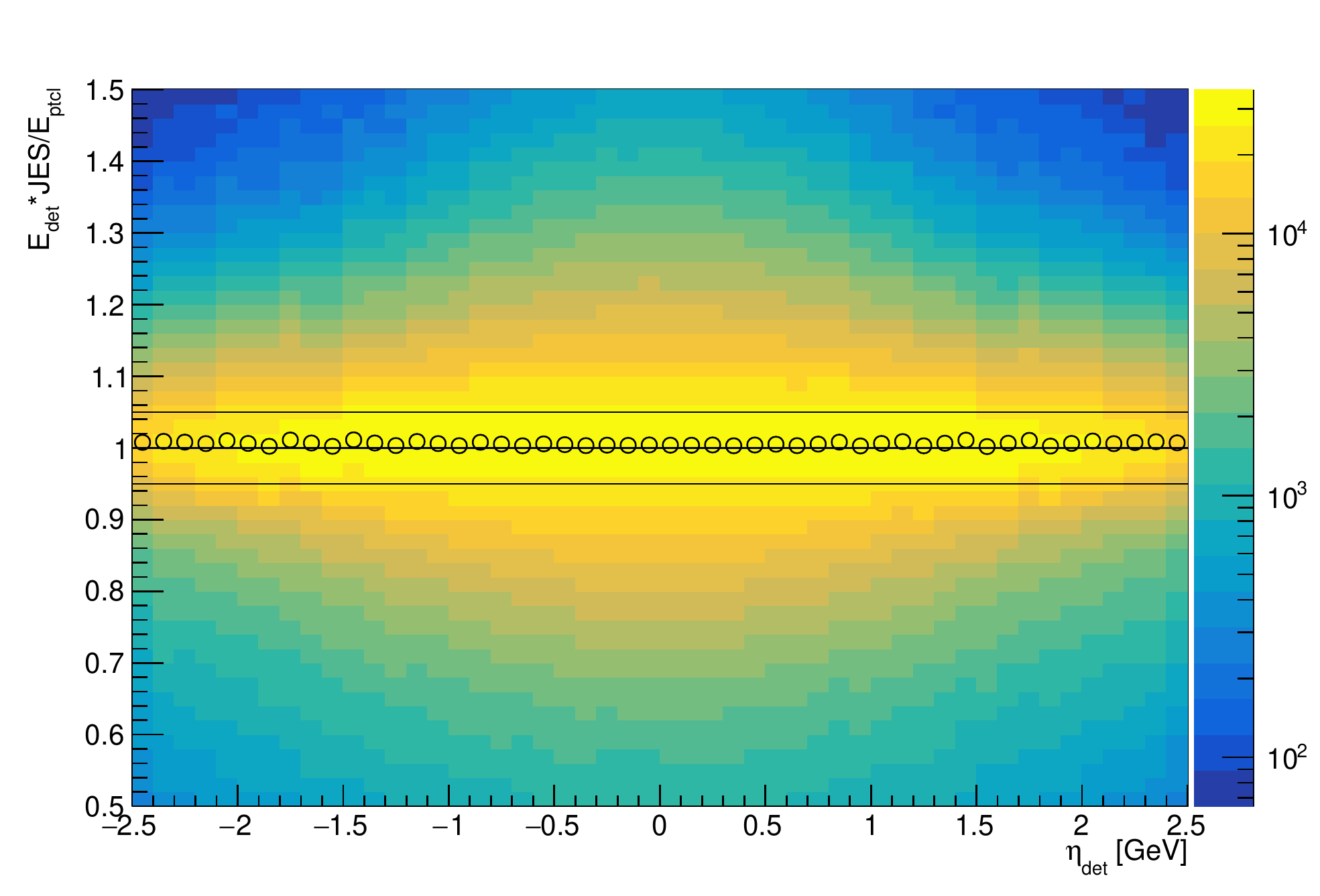} &
\includegraphics[width=0.45\textwidth]{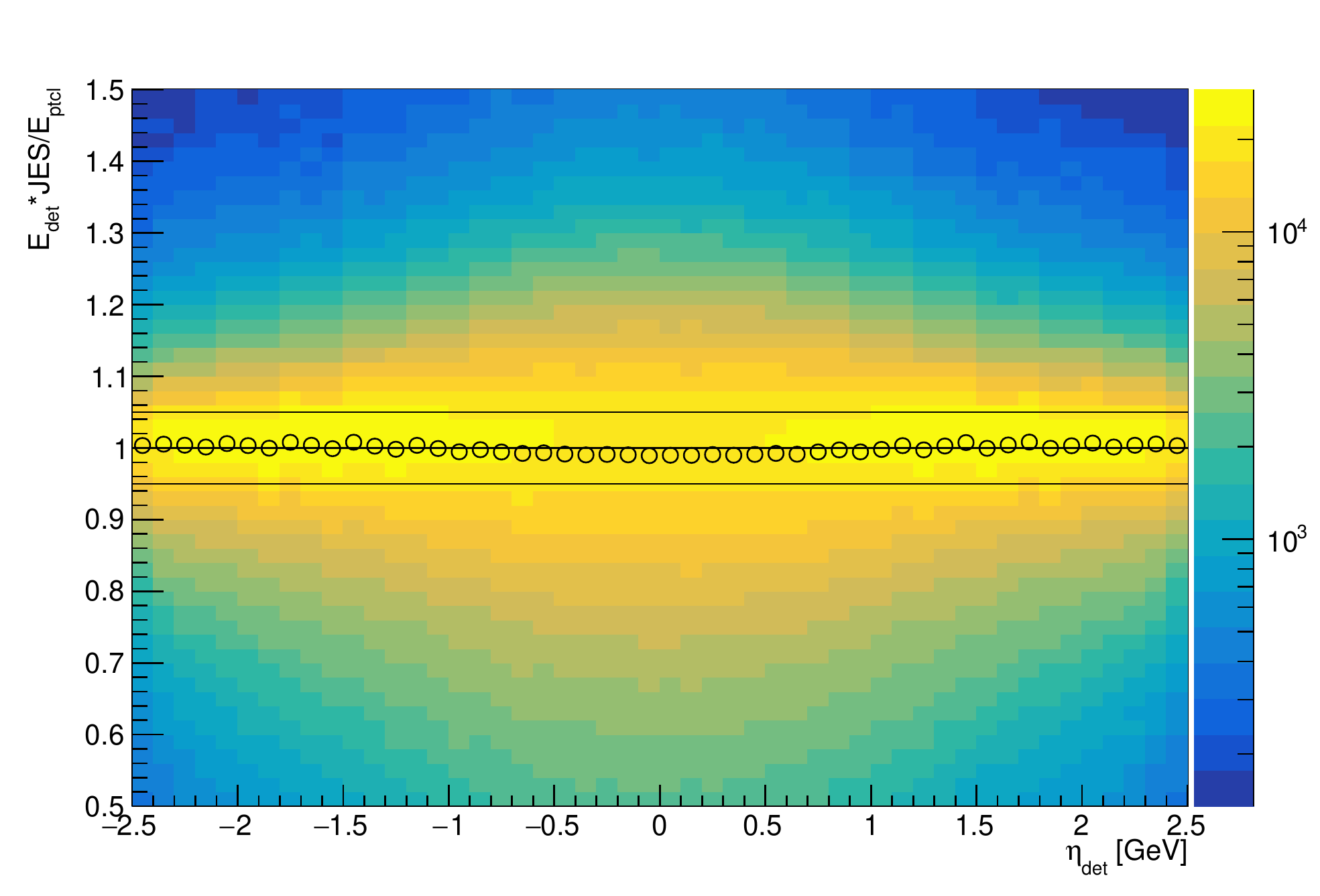} \\
\end{tabular}
\caption{Jet energy correction closure tests for the large jets (right) and small jets (left) as function of the transverse momentum (top) and $\eta$ (bottom) of jets.}
\label{fig_JES_clos}
\end{center}
\end{figure}

The closure test results as function of the transverse momentum and pseudorapidity $\eta$ of jets are shown for both small jets and large jets in Fig.~\ref{fig_JES_clos} where the mean deviation from unity is well within 5\%.  

\section{Top quark and $W$ boson tagging efficiencies}
\label{appendix_tagging}

The tagging of the large-$R$ jets originating from the top quark or the $W$ boson is a~common practice in high energy physics and was used also in this paper.
In order to evaluate the tagging efficiencies, a comparison to the generator level information is needed to define a truth jet label as top, $W$ or light otherwise.
The angularly closest large-$R$ jet to the direction of the original top quark was used as a probe for the truth top tagging efficiency $\varepsilon_\mathrm{top}$ which is defined as follows
\begin{equation}
\varepsilon_{\mathrm{top}} =\frac{N_{\mathrm{jet, top}}^{\mathrm{match\&tag}}}{N_{\mathrm{jet, top}}^{\mathrm{match}}}\,,
\end{equation}
where $N_{\mathrm{jet, top}}^{\mathrm{match\&tag}}$ is the number of jets matched to the original top quark and top-tagged by the tagging technique; and $N_{\mathrm{jet, top}}^{\mathrm{match}}$ is number of all jets tagged as possibly originating from top quark, as marked by the algorithm. The truth $W$ boson tagging efficiency is defined in a similar manner.
The mistag (fake) efficiency was also evaluated, which describes the false positivity of the tagger on jets not originating from the top quark or the $W$ boson. It is defined by the following formula in case of the top tagging
\begin{equation}
\varepsilon_{\mathrm{mis,top}} =\frac{N_{\mathrm{jet, top}}^{\mathrm{\neg match\&tag}}}{N_{\mathrm{jet, top}}^{\mathrm{\neg match}}}\,,
\end{equation}
where $\varepsilon_{\mathrm{mis,top}}$ is the mistag efficiency of the top quark tagger, $N_{\mathrm{jet, top}}^{\mathrm{\neg match\&tag}}$ is the number of large jets which are not matched to the generator level top quark but are top-tagged by the tagger; and $N_{\mathrm{jet, top}}^{\mathrm{\neg match}}$ is the number of all large jets not matched to the generator level top quark. The $W$ mistag efficiency is calculated in a similar way for the both detector and particle levels.

\begin{figure}
\begin{center}
\begin{tabular}{cc}
  \includegraphics[width=0.45\textwidth]{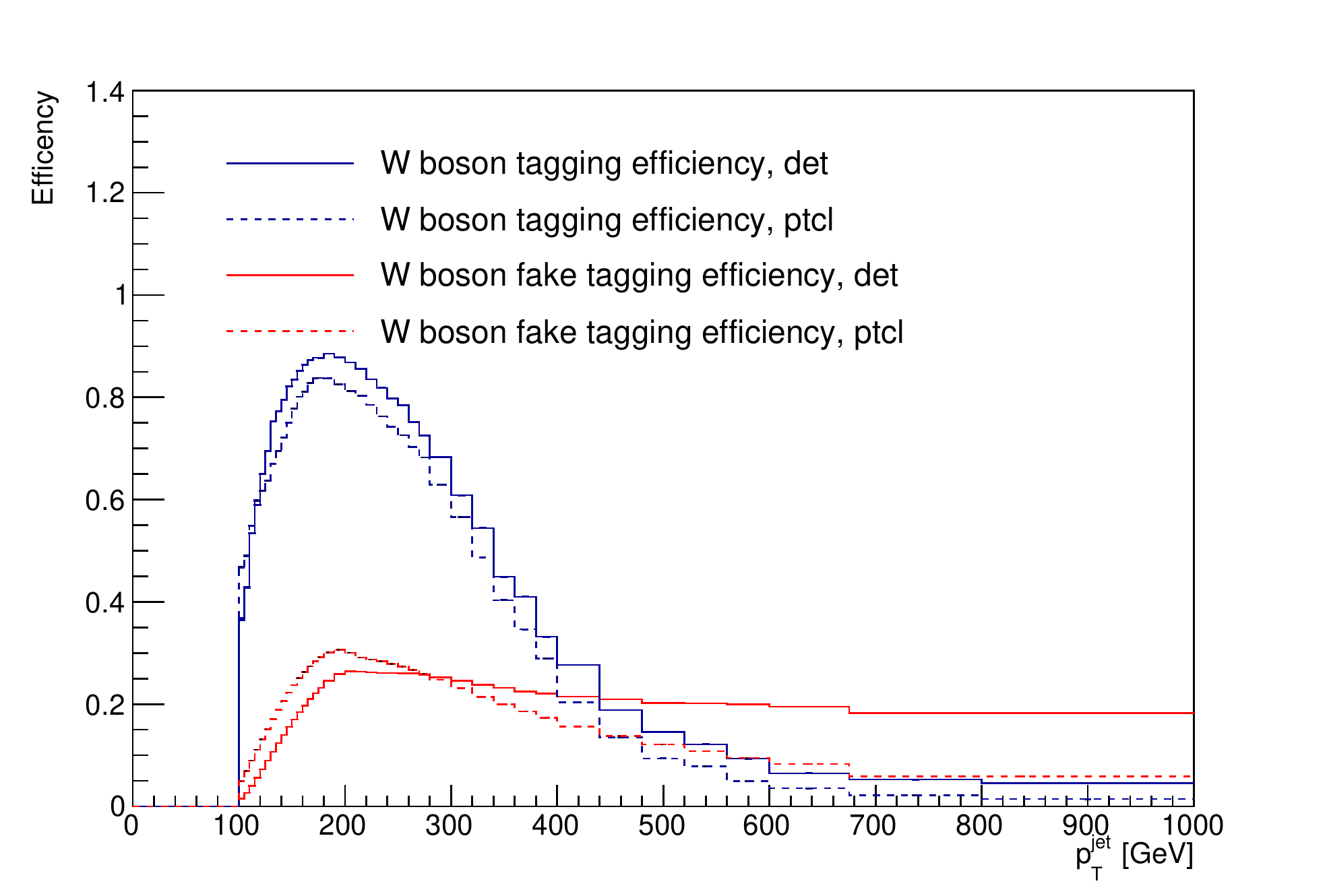} &
  \includegraphics[width=0.45\textwidth]{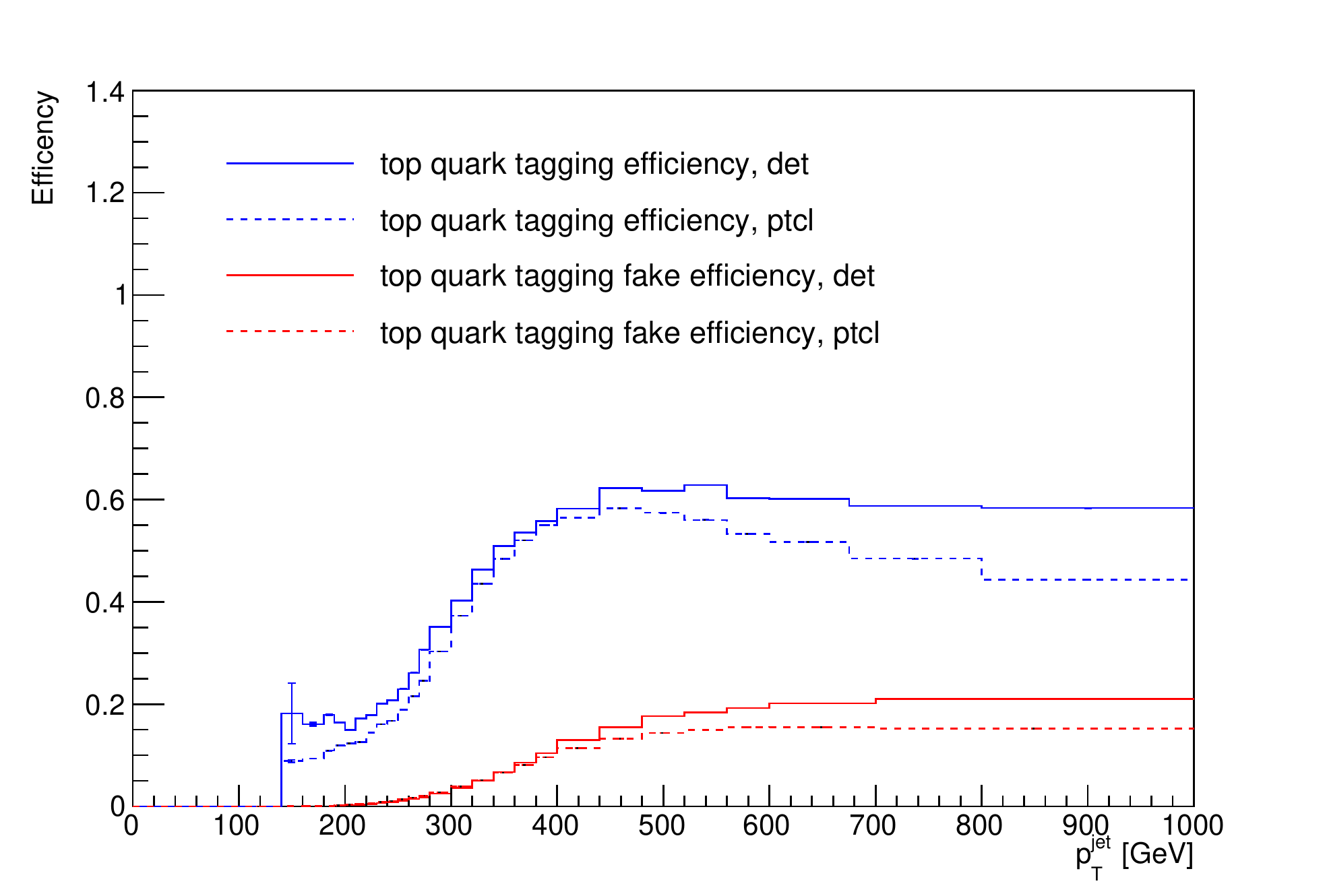} \\
\end{tabular}
\caption{Tagging (blue) and mistag (red) efficiencies for the $W$ boson (left) and for the top quark (right) in dependence on the transverse momentum of the large jet studied on the $t\bar{t}$ sample at both particle and detector levels.}
\label{fig_tag_effi}
\end{center}
\end{figure}

Both tag and mistag efficiencies are shown in Fig~\ref{fig_tag_effi} for the $W$ (left) and top (right) tagger, studied at both detector and particle levels in dependence on the transverse momentum of the large jet.
The sample for the mistag efficiency was the production of $2j2b$ events and contained 32.5M events, with events generated in exclusive $\pt$ ranges of the leading and sub-leading jets in order to populate the phase space of higher transverse momenta.

\end{appendices}
\end{document}